\documentclass[aps,pre,twocolumn,groupedaddress]{revtex4}
\usepackage{epsfig}
\usepackage{graphicx,color}
\usepackage{dcolumn}
\usepackage{natbib}
\usepackage{amssymb}

\def\dj{\hbox{d\kern-0,347em \vrule width0,3em height1,252ex
depth-1,21ex \kern0,051em}}

\begin{document}


\title{Rheology of colloidal microphases in a model with competing interactions}



\author{Alessandra Imperio$^1$
}
\author{Luciano Reatto$^2$}
 \affiliation{$^1$ CNISM, Sezione dell'Universit\`a degli Studi di Milano, Italy.\\
$^2$ Dipartimento di Fisica, Universit\`a degli Studi di Milano,via Celoria 16, 20133 Milano, Italy
}
\author{Stefano Zapperi$^{3,4}$}
\affiliation{$^3$ INFM-CNR, S3, Dipartimento di Fisica,
        Universit\`a di Modena e Reggio Emilia,
        via Campi 213/A, I-41100, Modena, Italy}
\affiliation{$^4$ ISI Foundation Viale San Severo 65, 10133 Torino, Italy}


\begin{abstract}

We study the rheological properties of colloidal microphases in two 
dimensions simulating a model of colloidal particles with competing 
interactions. Due to the competition between short-range attraction and 
long-range repulsion, as a function of the density the model exhibits a 
variety of microphases such as clusters, stripes or crystals with 
bubbles. We prepare the system in a confined microphase employing 
Monte-Carlo simulations and then quench the system at $T=0$. The resulting 
configurations are then sheared by applying a drag force profile. We 
integrate numerically the equation of motion for the particles and analyze 
the dynamics as a function of the density and the applied strain rate. We 
measure the stress-strain curves and characterize the yielding of the 
colloidal microphases. The results depend on the type of microphase: (i) 
clusters are easily sheared along layers and the relative motion is 
assisted by rotations. (ii) Stripes shear easily when they are parallel to 
the flow and tend to jam when are perpendicular to it. Under a 
sufficiently strong shear rate perpendicular stripes orient in the flow 
direction. (iii) Crystals with bubbles yield by fracturing along the 
bubbles and eventually forming stripes. We discuss the role of 
dislocations, emitted by the bubbles, in the yielding process.

\end{abstract}


\maketitle

\section{Introduction}
Examples of spontaneous pattern formation or micro-separation, in condition close to the thermodynamic equilibrium, can be found in many different experimental systems, both in two dimensions \cite{klokkenburg06,gelbart99,elias97,ghezzi97,seul95} and in three dimensions \cite{bardi07,lu06,stradner04, islam03}. The interest in these novel phases of matter is very high because the skill in controlling the architecture of particle aggregates, as well as the super-structures the latter arrange on, is nowadays a strategic tool in nanotechnology, in order to engineer new materials with specific electric, magnetic, optical and rheological properties. 

The topic of pattern formation is important not only from a technological point of view, but also from a fundamental one, influencing many different branches of science: from vitrification and gelation\cite{sciortino04,charbonneau06,decandia06,hoare07} to colloidal systems\cite{hecht07,reynaert06,puertas05,campbell04}, from biological membranes \cite{destainville06,goldman05} to network formation of tissue cells \cite{szabo07}, to  physisorbed layers on solid surfaces \cite{kern91,muller95}. Even if all of these systems are rather different at the molecular level, the patterns formed by the particles can be similar: circular clusters, stripes  and rings of particles can be easily observed in most of the bidimensional systems, while lamellae, spherical clusters and cylinders  occur in three dimensions. Moreover the particle aggregates can be often arranged to form disordered liquid-like configurations, but also crystalline-like super-structures such as grids of stripes and lamellae, hexagonal lattice of clusters and bubbles.

In many cases a simple mechanism to explain the micro-separation is the competition between a short-range attraction and a long-range repulsion. For colloidal systems, the short-range attraction can be due to depletion forces and Van der Waals forces, while the longer-range repulsion stems from dipolar forces as in Langmuir monolayers and ferrofluids, or from partially screened electrostatic forces. For sufficiently strong long-range repulsion, the standard vapor to liquid transition is inhibited in favor of the microseparation. In particular, the short-range attraction favors the condensation of particles, while the growth of the clusters is limited by the long-range repulsion. Recently, however, models based on purely repulsive interactions have been discussed which also support the micro-separations \cite{mladek06,glaser06,malescio03}.

Although great attention has been given to the study of the micro-separated systems via theoretical \cite{barci07,tarzia07,sciortino04,pini00} and numerical approaches \cite{reichhardt04,reichhardt03,imperio06,stoycheva02,larson92}, a general comprehension of the behavior of such phases is still lacking. Recently much attention has been also given to the role of confinement on pattern formation:
 simulations and experiments suggest, in fact, that confinement is a powerful tool to modify and even induce new pattern morphologies, which in the bulk might not appear at all. Most of these studies concern 3D block copolymer systems between smooth or patterned substrates \cite{tsori01,tsori06,li06,duchs04} and recently 2D colloidal systems under confinement \cite{impe07}.

Many common complex fluids such as pastes, emulsions, foams, colloidal systems and granular materials are studied also from the point view of the mechanical and rheological properties, which are relevant for industry. These systems, in fact, might exhibit both solid-like and liquid-like responses depending on the length scale, the time scale and on the applied shear stress. Moreover the shear has proved to be a mechanism capable of inducing and inhibiting pattern formation as well.  Many experiments involve aqueous foams under shear, which act as elastic solids for small deformations, while flow for large applied shear \cite{gopal95,gopal05}. Moreover cluster phases in colloidal systems, which become dilatant under flow, have been recently reported in \cite{lootens05,osuji07}. Finally in Ref.~\cite{cohen05} the role of geometrical confinement in the shearing of colloidal systems is considered.

Simulations in 3D have pointed out that shear can favor or inhibit the cluster phase with respect to the homogeneous one, via shearing \cite{hecht07}. In particular in Ref.~\cite{melrose92}, the authors have analyzed the behavior of mixtures of particles both in the brownian regime, where the shear forces are comparable to the brownian ones, and in the Stokes regime where the shear forces greatly exceed the brownian forces. In both cases, systems which form gels at rest, support the formation of much more compact and somehow complex structures under shear. The role of external stresses, close to the jamming transition and during the aggregation processes in attractive colloidal systems, is presented in \cite{trappe01}, where a general phase diagram is proposed as a function of density, temperature and stress.

Concerning rheological properties of 2D systems, most of the experiments focus on the homogeneous phases of Langmuir monolayers \cite{walder07, ignes-mullol07}. Recently attention has been paid also to the inhomogeneous phases as in Ref. \cite{lauridsen05}, where a single layer of bubbles floating on a liquid surface is studied. Very interestingly such a system shows the coexistence between a flowing and a jammed phase of bubbles. Simulations instead address mainly the properties of foams as in \cite{durian99,tewari99,okuzono94}, where flow is strongly intermittent, once the shear stress exceeds the yield stress. The avalanche-like rearrangement of the bubbles 
are visible from the stress-strain curve, where the fluctuations are strongly enhanced. In other words the stress is at first accumulated by macroscopic deformation of the structure and then abruptly released, followed by topological changes. Stain bursts are typical in the plastic deformation of amorphous solids, as shown in 
numerical simulations \cite{maloney04,demkowicz05,bailey07}, but also of microscale crystals \cite{uchic04,dimiduk06} 
where they are due to the collective dynamics of dislocations \cite{miguel01,csikor07}.

In this paper we present results of numerical simulations on the rheology of two dimensional colloidal microphases in confined geometries. We consider an ensemble
of colloidal particles interacting by a potential with short-range attraction
and long-range repulsion. As a function of the density and depending 
on the constraints imposed by the confinement, the colloidal system displays a variety of equilibrium microphases, such as clusters, stripes or bubbles \cite{impe07}. 
Here, we analyze the behavior of the system when a shear profile is imposed
on the fluid. In particular, we focus on the stability of the pattern morphology under shear, especially in the intermediate high-density region which has not been studied so far. The rheological behavior of the system is dependent on the
density and on the corresponding microphase. At low densities, the colloidal
particles separate into clusters that are easily sheared. In this case we observe
that shearing is assisted by cluster rotations, a process that is
reminiscent of the shear of granular media \cite{astrom00}. At higher densities
the colloidal particles form stripes. When the stripes are oriented parallel
to the shear direction, the system deforms easily, following the imposed 
velocity profile. On the other hand, when the stripes are oriented perpendicular
to the shear direction the system tends to jam. Only under the application of  sufficiently high shear rate the system is able to flow. This is done by a reorientation of the stripes along the shear direction. Similarly, bubble
phases are jammed at low shear rates and flow at large shear rate. In order
to flow, the bubbles tend to coalesce into stripes which are then easily 
sheared. 

The aforementioned rheological behavior is reflected in the stress
strain characteristics of the system. In the case of perpendicular stripes
and bubbles, the shear stress displays a yield point, corresponding
to the break-up of the jammed configurations, followed by a shear
weakening regime. For large shear strain, the stress reaches a constant value
corresponding to the flow of parallel stripes.
Shearing a system which is prone to microseparate is similar to the
application an external force field which favors specific pattern morphologies or specific orientations. The possibility to control, modify and stabilize a specific pattern is particularly appealing for the technology of new materials. From this point of view, the application of shear stress could represent a promising strategy to 
control the structure of microphases.

This paper is organized as follows. First we describe the microscopic model of the system and its equilibrium properties, in the bulk and under confinement, in section \ref{equil}; then the rheological quantities as well as the simulations details are introduced in section  \ref{rheology}. Then in sections  \ref{Cluster}, \ref{Stripe} and \ref{Bubble}, respectively, we describe the rheological properties for the cluster, the stripe and the bubble phases. In particular, for each case we discuss the shear rate profile as a function of time, the flow curve (that is the stress-strain curve) and the
 velocity profile of the particle aggregate as a function of the distance from the confining walls. A comparison between the behavior of the fluid under shear, adopting different pattern morphology, is drawn in section  \ref{Discussion}.

\section{The model \label{The model}}

\subsection{\label{equil}Equilibrium }
Within the framework of the competing interactions, we consider a bidimensional fluid, in which the effective potential for a pair of colloidal particles is:
\begin{widetext}
\begin{equation}
U_{pp}(r)= \left\{\begin{array}{ll}
                                A\left(\frac{\sigma}{r}\right)^n-\frac{\epsilon_a\sigma^2}{R_a^2}\exp(-\frac{r}{R_a}) +\frac{\epsilon_r\sigma^2}{R_r^2}\exp(-\frac{r}{R_r}) & \mbox{if $ r\leq R_{cut}$}\\
                                   0 & \mbox{otherwise}\\
       \end{array}
\right.
\label{Upp}
\end{equation}
\end{widetext}

\begin{figure}
\includegraphics[width=8cm]{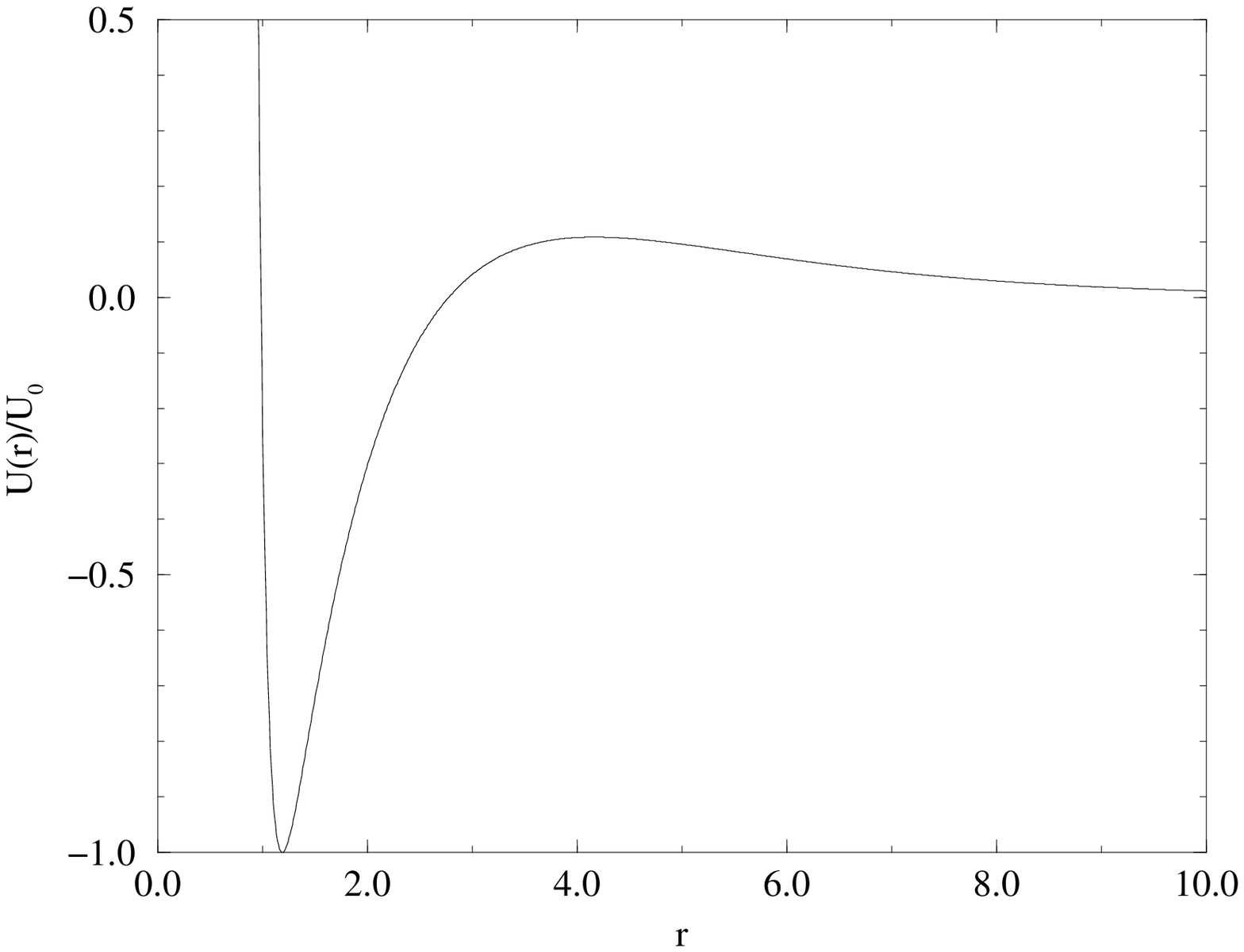}\\
\includegraphics[width=8cm]{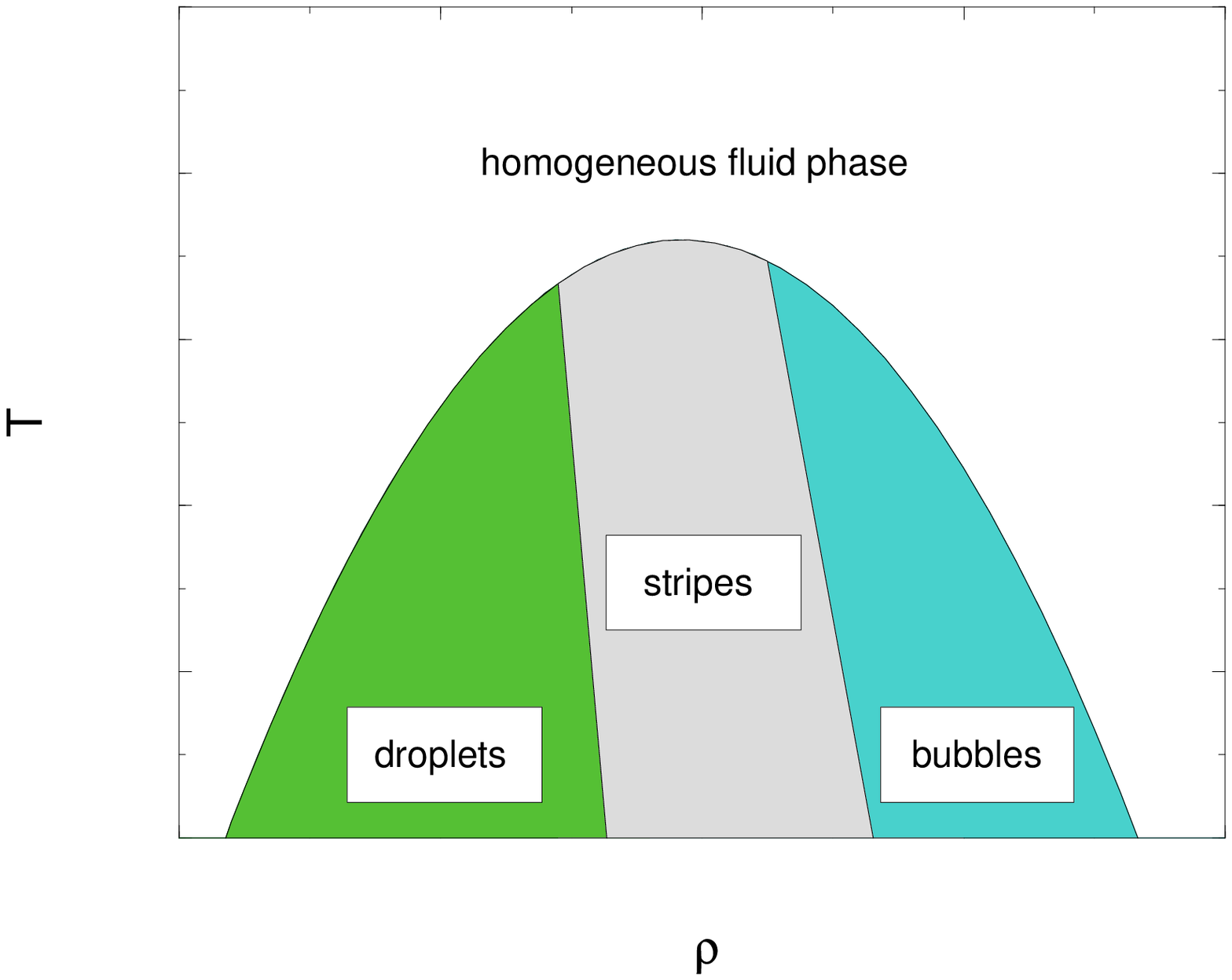}
\caption{\label{phase_diagram}\footnotesize{Upper panel: particle-particle effective potential corresponding to Eq.~(\ref{Upp}, in units of the potential well depth $U_0=|U_{pp}(r=\sigma)|$). Bottom panel: schematic phase diagram, in the density, temperature plane ($\rho,T$) for a bulk bidimensional fluid subject to competing interactions.} }
\end{figure}

  $r$ being the interparticle distance and $\sigma$ the particle diameter. Such a potential is characterized by a soft-core repulsive term at very short distances, followed by an attractive well plus a repulsive hump at larger distances (see Fig.~\ref{phase_diagram}, upper panel). The potential parameters are: $n=12$, $A=0.018$ for the short-range repulsion, $R_a~=~1~\sigma,\epsilon_a~=~1$ for the short-range attraction and $ R_r~=~2~\sigma, \epsilon_r~=~1$ for the longer-range repulsion. The short-range repulsion takes into account the impenetrability of the particles, while the remaining part of the potential mimicks the colloid interaction mediated by the surrounding fluid.  The potential is truncated at $R_{cut}~=~10~\sigma$. With the present choice of the parameters, the minimum of the potential is located at $r\sim1.2\sigma$, while the maximum value of the longer-range repulsive hump is at $r\sim4\sigma$ .\\

 The interaction potential of equation (\ref{Upp}) is similar to those studied in \cite{tarzia07,decandia06, sciortino04} for three dimensional systems. In case of two-dimensional fluids our model is the same discussed in \cite{imperio06,impe07}, but for the short-range repulsion which is treated as a hard-core in the latter, while  it is implemented as a soft-core potential in the present work. 
Such a difference is not particularly important from the point of view of the general features of the phase diagram in bulk, which is schematically depicted in Fig.~\ref{phase_diagram} (bottom panel). At sufficiently low temperature, the present model supports the passage from droplets to stripes to bubbles as the density increases. The main difference with the bulk model studied in \cite{imperio06} is merely an overall shift of the phase diagram towards lower densities.

 In our model, the fluid is confined along the $x$ direction by smooth parallel walls, the separation of which is $L_x$. The wall-particle interaction is treated as a rather steep potential of the form
\begin{equation}
U_{wp}(x)=  C \exp(-x/\xi),
\label{Uwp}
\end{equation}
where $x$ is the distance from the wall. The parameters we use in simulations are $\xi=0.1\sigma$
and $C=U_0$, $U_0$ being the depth of the attractive well of the particle-particle potential. As the range of the wall-particle interaction is very short and the repulsion very strong, such a potential mimicks very well the effects of neutral hard walls, as we have also verified performing some Monte Carlo simulations with a hard-wall potential too.

Under lateral confinement, if the hard disk potential is used for the short-range repulsion as in \cite{impe07}, a transition from stripe to clusters occurs as the temperature is lowered. Such a transition disappears in our case, where the soft-core potential repulsion is used. The results discussed in  \cite{impe07} can be obtained if the exponent $n$ of Eq.~(\ref{Upp}) is increased up to 36. But as the aim of this work is studying the effect of shear on different pattern morphologies, disregarding at first the thermal noise, the change of morphology with temperature is not of interest at this stage. Moreover, with the present choice of the potential parameters, we are able to study the stripe and bubble phase at relatively lower densities, meaning that we carry out simulations with smaller systems.

In this work we discuss three densities $\rho\sigma^2~=~0.15$,~$0.3$,~$0.5$, corresponding to the cluster, stripe and bubble regime.  The clusters and the bubbles are arranged onto a triangular super-lattice, so that the particle configurations are specular. One of the goals of our work, in fact, is to study how such a hypothetical symmetry between filled and empty region in the particle configuration affect the rheology of the fluid. The stripes, instead,  form a grid of parallel layers. In each microphase the pattern period is $P~\sim~11~\sigma$. The crystalline order of the super-lattice is signaled by the occurrence of Bragg peaks in the static structure factor at short wave vectors whose length is roughly connected to the period through the relation $P\approx 2\pi/K_p$. Random phase approximation for this kind of potential has provided predictions in good agreement with simulations \cite{imperio06}. The maximum size for the particles aggregates, instead, is dominated by the width of the attractive well of the particle-particle potential. So that, in the present case, we see that cluster diameter is $\sim~4~\sigma$, similarly to the bubble diameter and to the stripe width.

\begin{figure}
\includegraphics[width=4cm,height=4cm]{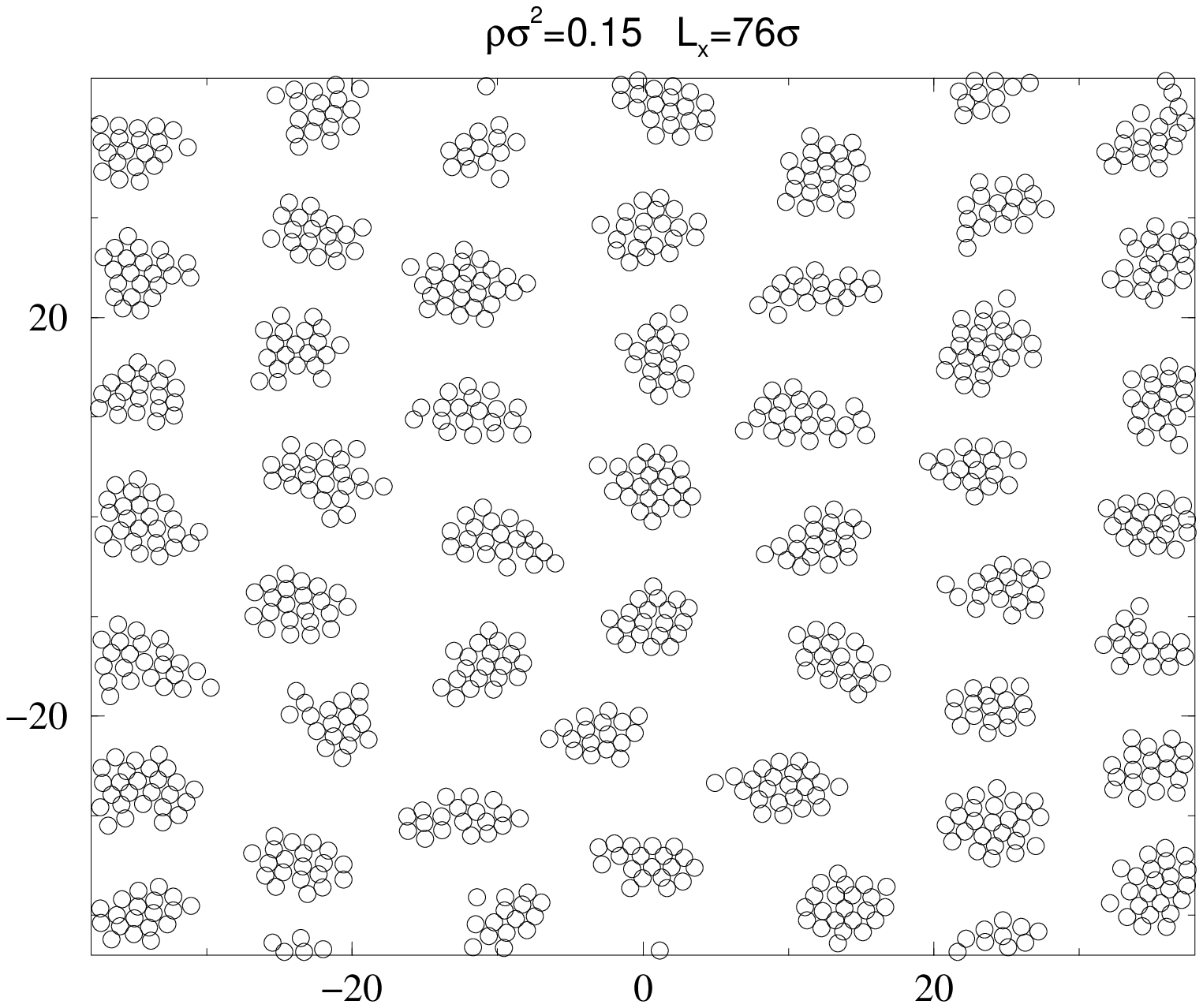}\includegraphics[width=4cm,height=4cm]{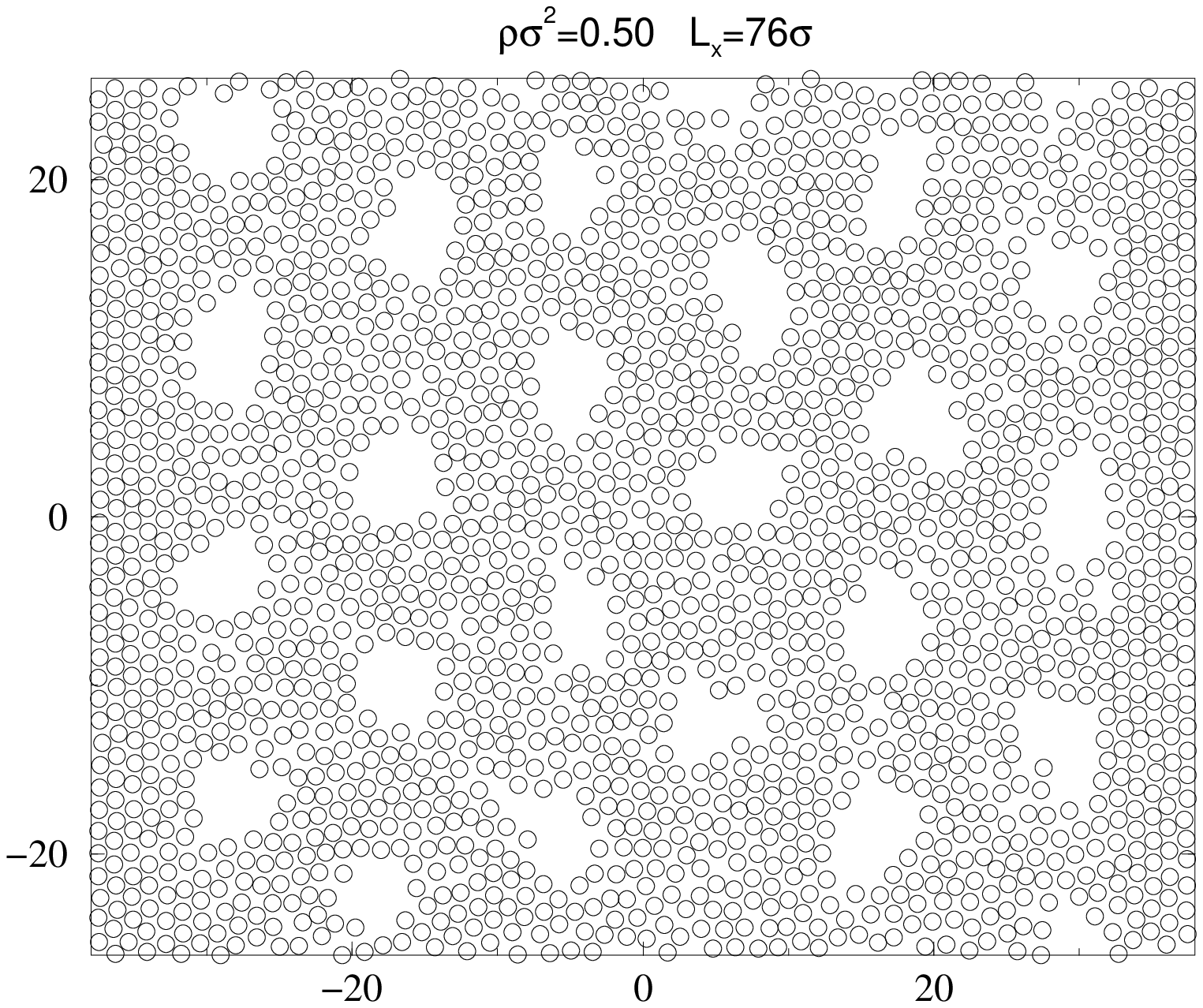}\\
\includegraphics[width=4cm,height=4cm]{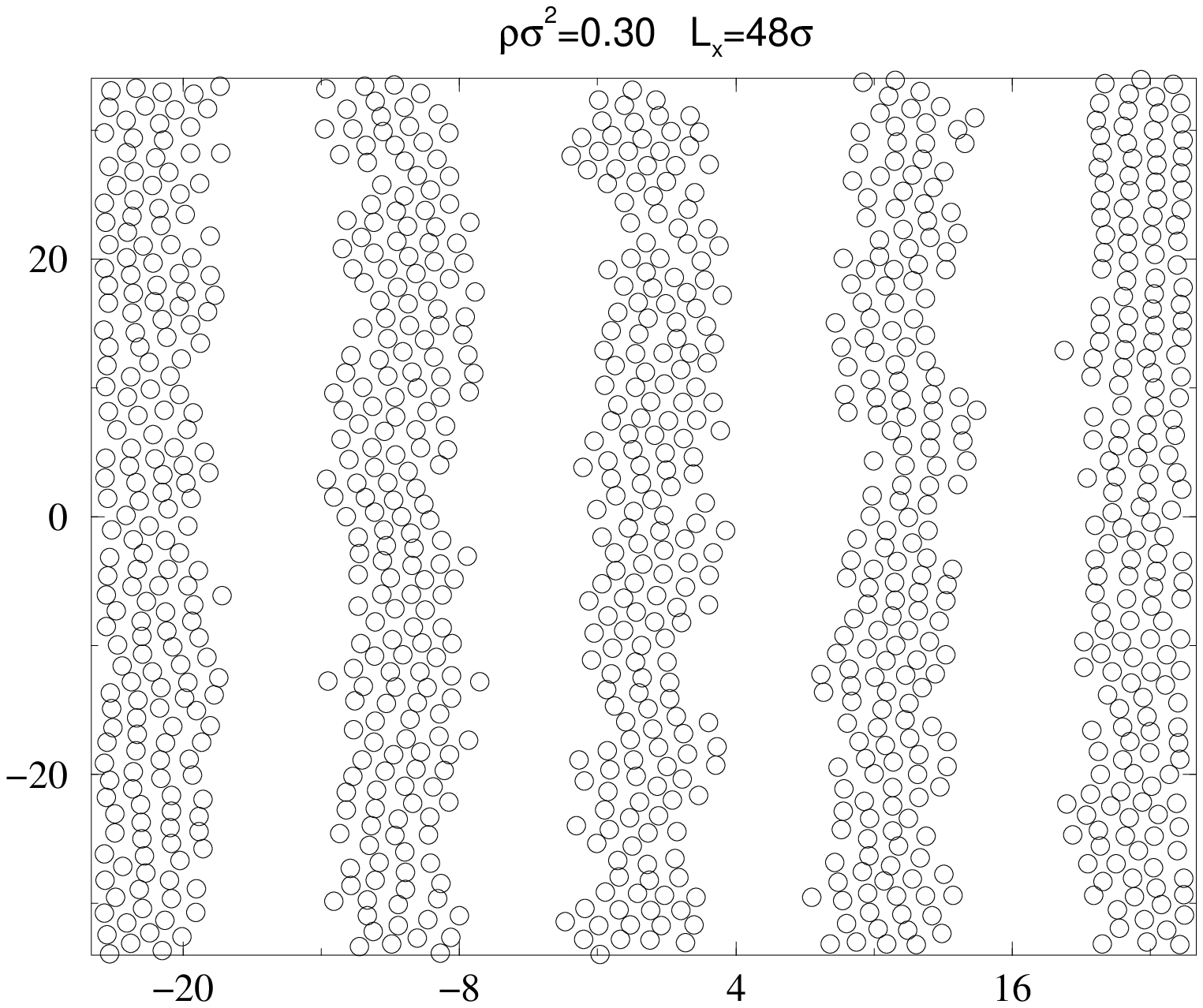}\includegraphics[width=4cm,height=4cm]{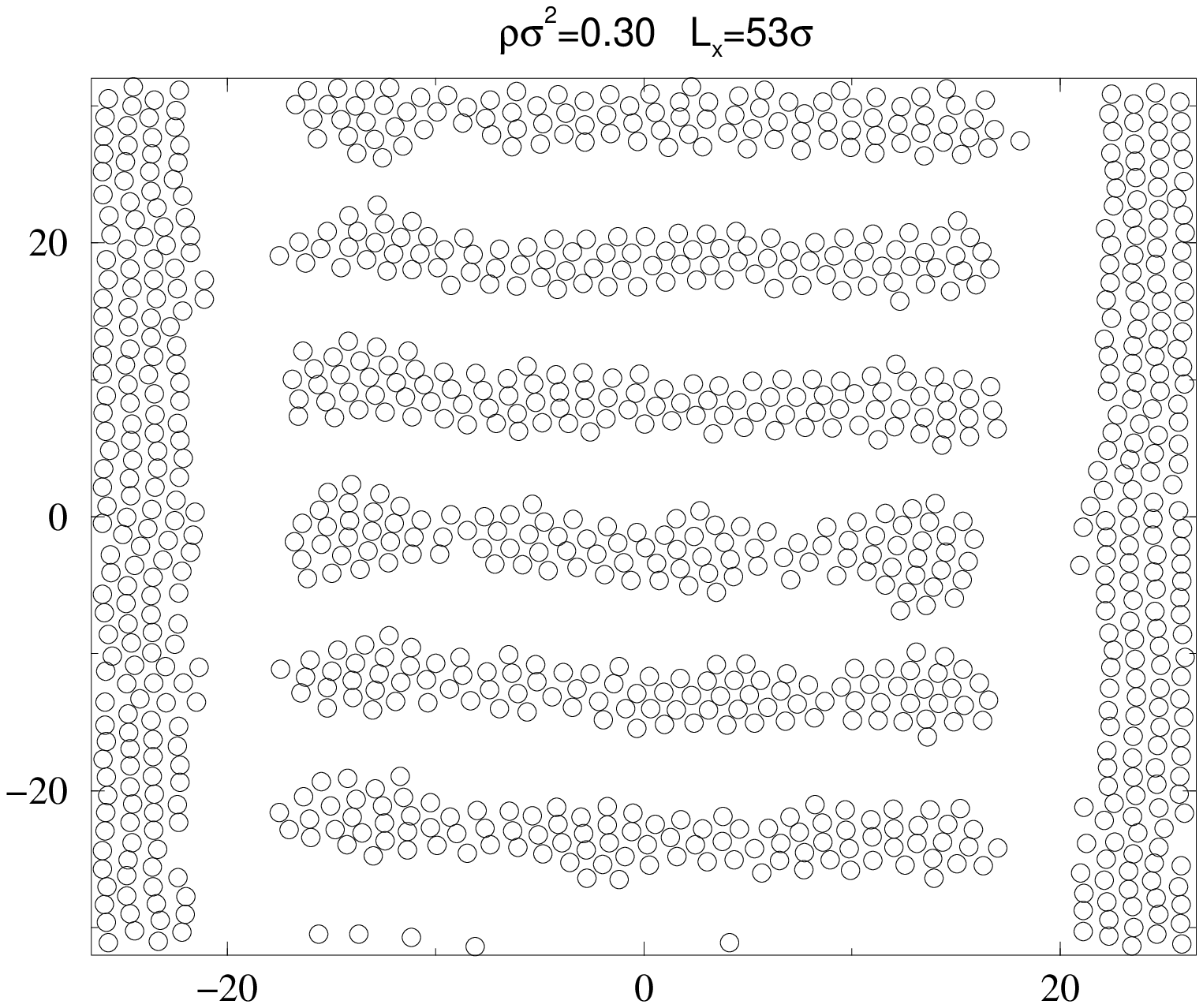}\\
\caption{\label{snap_equilibrium}\footnotesize{Typical configurations obtained via Monte Carlo simulations for different densities and different values of the wall separation $L_x$ and $T=0.4U_0$. We notice that when the wall separation $L_x$ is not commensurate to the stripe period, the perpendicular alignment is favored upon the parallel one.} }
\end{figure}

The addition of a geometrical constraint, such as the presence of neutral walls along one of the directions, is a source of further frustration for the system, mainly if the wall separation $L_x$ is not commensurate to the intrinsic bulk periodicity. For our model, the most peculiar case is represented by the striped phase, where the alignment of the stripes  depends strongly on the value of $L_x$. In particular, if $L_x= m P + \Delta_s$ ($m$ is an integer and $\Delta_s$ the stripe width), the stripe orientation is parallel to the walls, otherwise we observe the formation of a mixed phase with two parallel stripes next to the walls and stripes perpendicular to the walls in the center of the slit. In figure \ref{snap_equilibrium}, typical configurations for different patterns between walls are plotted. Such configurations are generated through Monte Carlo simulations in the canonical ensemble ($NVT$, $N$ number of particles, $V$ volume, $T$ temperature). In our simulations $N$ ranges from 1000 to 2000 particles, and the temperature is equal to $T=0.4 U_0$.

\subsection{Rheology \label{rheology}}

To simulate the rheology of 
the colloidal system, we consider a simple overdamped
dynamics for the particles under the action of an external force field due to
the shear. A moving colloidal particle in a fluid is subject to a drag force.
In particular, for a spherical particle of radius $a$ moving at small
velocity $v$, the drag force is given by $F_d= -\Gamma v$, with $\Gamma=6\pi\eta a$
where $\eta$ is the fluid viscosity. In this paper, we apply a weak linear shear 
profile to the carrier fluid $\vec{V}(x)= V_0 x \hat{y} /L_x$ and, correspondingly,
the colloid particle is subject to a drag force profile $\vec{F}_d(x)=F_0 x \hat{y}/L_x$, where $F_0=\Gamma V_0$. Neglecting inertial and thermal effects, the equation of motion for the particles is given by:
\begin{equation}
\Gamma\frac{d \vec{r_i} }{dt} =  \vec{F}_d(x)+\sum_j  \vec{f}_{pp}( \vec{r}_j-\vec{r}_i),
\label{eqm}
\end{equation}
where the interparticle forces are computed from the potential in Eq.~(\ref{Upp})
\begin{equation}
\vec{f}_{pp}(\vec{r})= - \frac{\partial U_{pp} }{\partial \vec{r}}.
\end{equation}
Eq.~(\ref{eqm}) is integrated numerically by means of an adaptive step-size 
fourth-order Runge-Kutta algorithm. In the present simulations, we fix the
timescales setting $\Gamma=1$. Equilibrium configurations
obtained from Monte-Carlo simulations at $T=0.4 U_0$ are first relaxed at
$T=0$ in absence of external forces, integrating Eq.~(\ref{eqm}) for $F_0=0$. 
Then we apply an external shear profile to the resulting configurations and
study the system as a function of $F_0$ and $\rho$.

It will be instructive to compare the internal shear strain rate  with the externally imposed strain rate $\dot\gamma_{ext}\equiv \frac{1}{2}\frac{\partial V_y}{\partial x} = V_0/(2L_x)$. The internal strain rate is in general not uniform and it is necessary to consider 
its average value defined as
\begin{equation}
\langle \dot\gamma\rangle= \frac{2}{L_x}\left[<v_y^+>-<v_y^->\right],
\end{equation}
where $<v_y^{\pm}>$ is the $y$ component of the particle velocity restricted to the $x\gtrless 0$ domain. To obtain a clearer picture of the internal 
strain rate, in order  to quantify the deviations from a laminar flow, 
we measure the particle velocity profile $\langle v_y(x) \rangle$ dividing the 
system into $n_s$ strips oriented along $y$ and averaging the $y$ component
of the velocities of the particles in each strip. In particular, we use $n_s=7$ 
for $L_x=76$ and $n_s=5$ for $L_x=48,53$. 
Finally, we measure the stress-strain curves of the system computing the
internal shear stress as
\begin{equation}
\sigma_{xy} = -\frac{1}{L_x L_y}\sum_{ij} (x_i-x_j) f_ {pp}^{(y)}(\vec{r}_i-\vec{r}_j).  
\end{equation}

\section{Cluster phase\label{Cluster}}

\begin{figure}[hbtp]
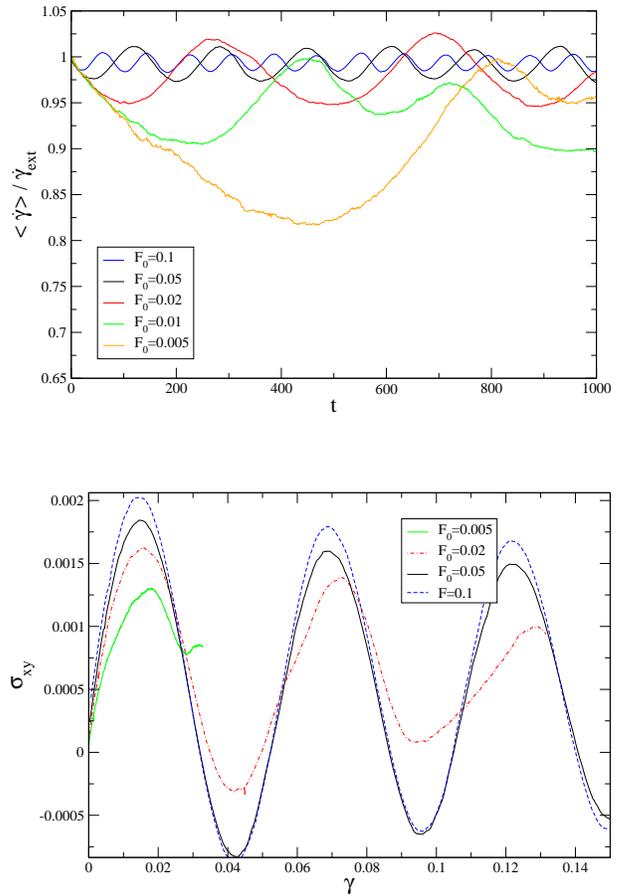

\includegraphics[width=8cm]{rho015gammadot.eps}\\
\vspace{1cm}
\includegraphics[width=8cm]{rho015stress-strain.eps}
\caption{(Color online)\footnotesize{ Upper panel: the average strain rate, divided by the external
strain rate $\dot\gamma_{ext}~\equiv ~V_0/~(2L_x)$, as a function of time for different values of $F_0$ for a system with $N=1000$, $\rho=0.15$ and $L_x =76$. Bottom panel: the stress-strain curves for the same cases shown in the upper panel.}}
\label{rho015gdot}
\end{figure}

The rheological properties of the cluster phase are ruled by the interactions
between individual clusters. At low temperatures individual particles inside the clusters
do not move, thus the clusters can be considered as rigid objects which repel each other. 
In Fig.~\ref{rho015gdot} (upper panel) we report the average strain rate $\dot\gamma$ of the system normalized by the external strain rate. The average strain rate shows characteristic oscillations, which are a peculiarity of the pattern geometry:  at low temperatures the clusters arrange into a regular super-lattice and to flow the system must overcome a Peierls-Nabarro periodic potential. In other words, the oscillations are connected to the sliding past of clusters moving along adjacent columns. The period of the oscillations diminishes as the external applied force $F_0$ increases, as the clusters move faster. Oscillations are also present in the stress-strain curves as shown in Fig.~\ref{rho015gdot} (bottom panel). For the timescales and shear rates that we have studied, we see a slowing
down of the dynamics at low $F_0$ but the system does not jam. We would
expect that at very low $F_0$ the system could be blocked by the weak
intra-cluster potential, but we could not simulate efficiently a similar
regime as the relaxation time becomes too long.

\vspace{0.5cm}

\begin{figure}[h!]
\includegraphics[width=8cm]{rho015prof.eps}\\
\vspace{1cm}
\includegraphics[width=8cm]{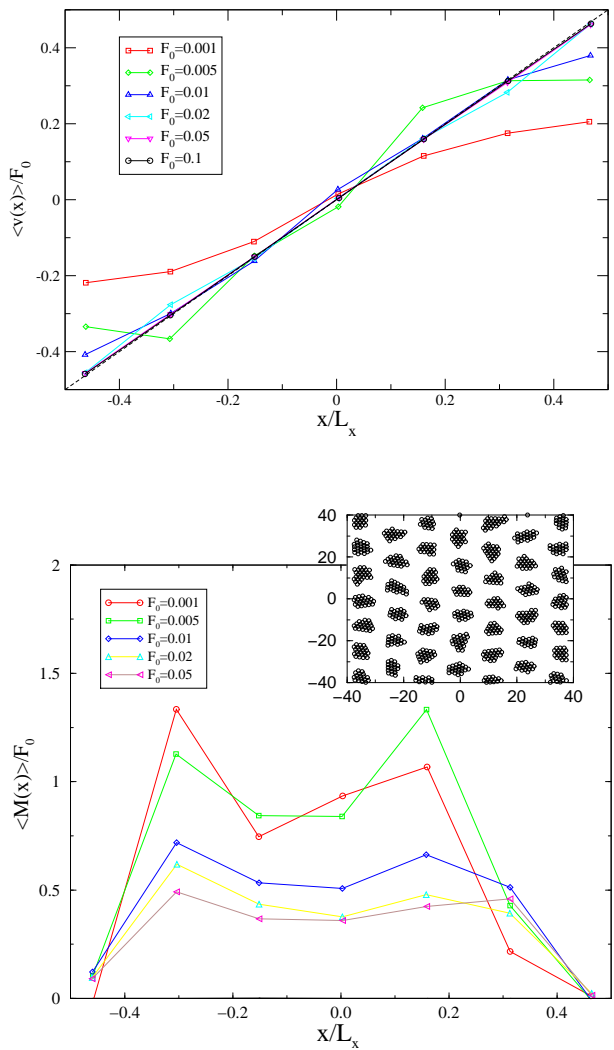}
\caption{(Color online)\footnotesize{ Upper panel: the velocity profiles for different values of $F_0$, for a system with $N=1000$, $\rho\sigma^2~=~0.15$ and $L_x~=~76\sigma$. Notice the deviation
from the externally imposed linear profile (dashed line) occurring at low shear rates. Data are averaged over time and over 10 realizations. Bottom panel: the torque $M$ profiles acting on the clusters. In the inset we show a typical configuration of the system.} }
\label{rho015prof}
\end{figure}

In  Fig.~\ref{rho015prof} (upper panel) we show the velocity profile of the clusters. The external applied velocity profile is also shown as a dashed line. At low $F_0$ the velocity profile shows that the colloid particles are slower with respect to the surrounding medium. Increasing the external force, instead, the colloid velocity profile tends to be equal to the external one (dashed line). In particular, as $F_0$ increases, the velocity profile tends to be linear at first in the middle of the box and then nearby the walls. Hence the flow is at first newtonian in the middle of the box, while slippage can be present close to the walls. Such a behavior can be considered as an indication of shear banding, where different spatial regions of the fluids are characterized by different flowing properties such as different values of the viscosity.\\
We see that the cluster lattice is distorted by the shear deformation
and clusters in different columns have to slide past each other. This process
is occasionally helped by rotations around the clusters center of mass.
Due to the shear profile individual clusters are subject to a net torque
that can induce rotations, creating a sort of roller-bearing effect
(See EPAPS Document No. XXXX.1 for an animation of the shear of the
cluster phase).
Similar rotations are commonly observed in the shear of granular media
and is very important in earthquakes when the faults are filled by a 
granular gouge \cite{astrom00}. Similarities between colloidal systems and granular media
have been pointed out on the basis of rheological measurements in Ref. \cite{lootens05}. 
Here, we measure the average torque of the clusters as a function of the
horizontal coordinate of their center of mass. As can be seen in 
Fig.~\ref{rho015prof} (bottom panel) the torque displays a characteristic pattern
depending on the cluster position. The clusters nearby the walls rotate less than the rest because the interaction with the surrounding clusters is limited on one side; the central clusters, instead, are those for which the applied velocity is less important as most of the particles experiment an almost zero drag velocity, so that their rotation is essentially due to the repulsion with the clusters flowing in the nearby columns. Finally the clusters whose position is intermediate between the walls and the center of the simulation box, experiment both the effects due to the applied drag force and to the cluster-cluster repulsion, so that the net torque is the highest.

\vspace{0.5cm}


\section{Stripe phases\label{Stripe}}

\begin{figure}[hbtp]
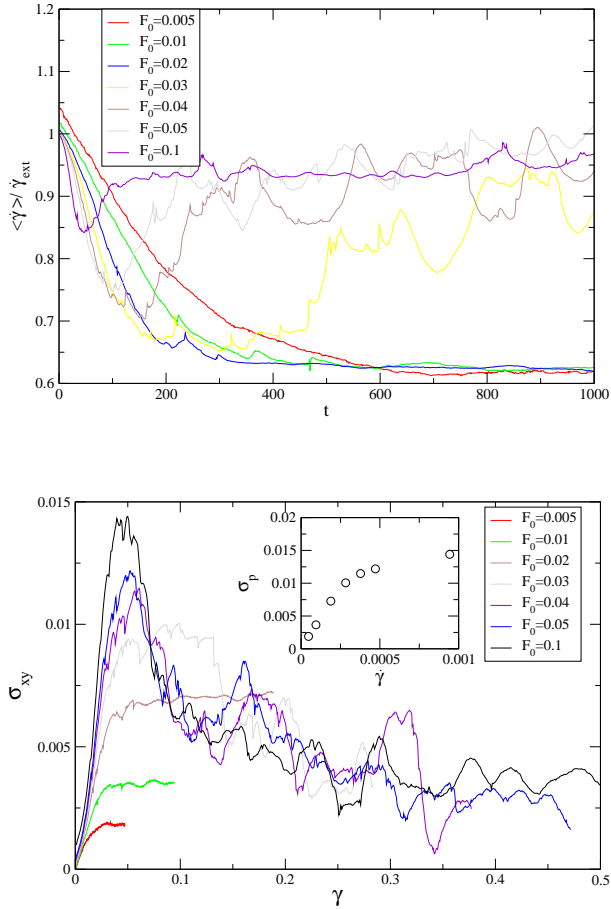

\includegraphics[width=8cm]{rho03_gammadot_per.eps}\\
\vspace{1cm}
\includegraphics[width=8cm]{rho03stress-strain_per.eps}
\caption{\footnotesize{(Color online) Upper panel: the average strain rate as a function of time for different
values of $F_0$ for a system with $N=1000$, $\rho=0.3$ and $L_x =53$ plotted in log-linear scale. In the initial condition the
stripes are mostly perpendicular to the flow direction. Bottom panel: the stress-strain curves. In the inset we report
the maximum stress $\sigma_p$ as a function of the strain rate $\dot\gamma$.
Notice the change of behavior as the systems yields at higher rates.}}
\label{rho03gdot_per}
\end{figure}

When we consider the rheology of the stripe phase, we have to pay attention
to the orientation of the stripes with respect to the shear direction. By
slightly tuning $L_x$ we can obtain configurations in which the stripes are all parallel to the walls or mixed configurations made up of two stripes parallel to the walls, while in the middle they arrange perpendicularly to the walls. \\ The most interesting case is that corresponding to the perpendicular stripes. In fact, at low forces the
system jams, failing to reorient in the flow direction. Hence, for $F_0 <0.03$, the average strain rate is lower than the external rate, while for larger $F_0$ their ratio tends to one (see Fig.~\ref{rho03gdot_per}, upper panel). This behavior is reflected by the stress strain curve (Fig.~\ref{rho03gdot_per}, bottom panel), that changes character for $F_0 \ge 0.03$ where the yield stress is followed by a strain-rate weakening part while,
for $F_0 <0.03$ the stress reaches a plateau. The peak stress for low $F_0$
scales in a viscous manner as $\sigma_p \simeq C\dot\gamma$ and changes
character for $F_0>0.03$ scaling as $\sigma_p \simeq \sigma_Y+C'\dot\gamma$,
with $C'<C$ (inset of Fig.~\ref{rho03gdot_per}). We also note strong fluctuations in the high strain regime, a behavior which is often observed in systems with avalanche-like flows as in \cite{maloney04,demkowicz05,bailey07,miguel01,csikor07,okuzono94,tsamados07}.

\begin{figure}[hbtp]
\includegraphics[width=8cm]{rho03prof_perp.eps}\\
\includegraphics[width=3cm,angle=-90]{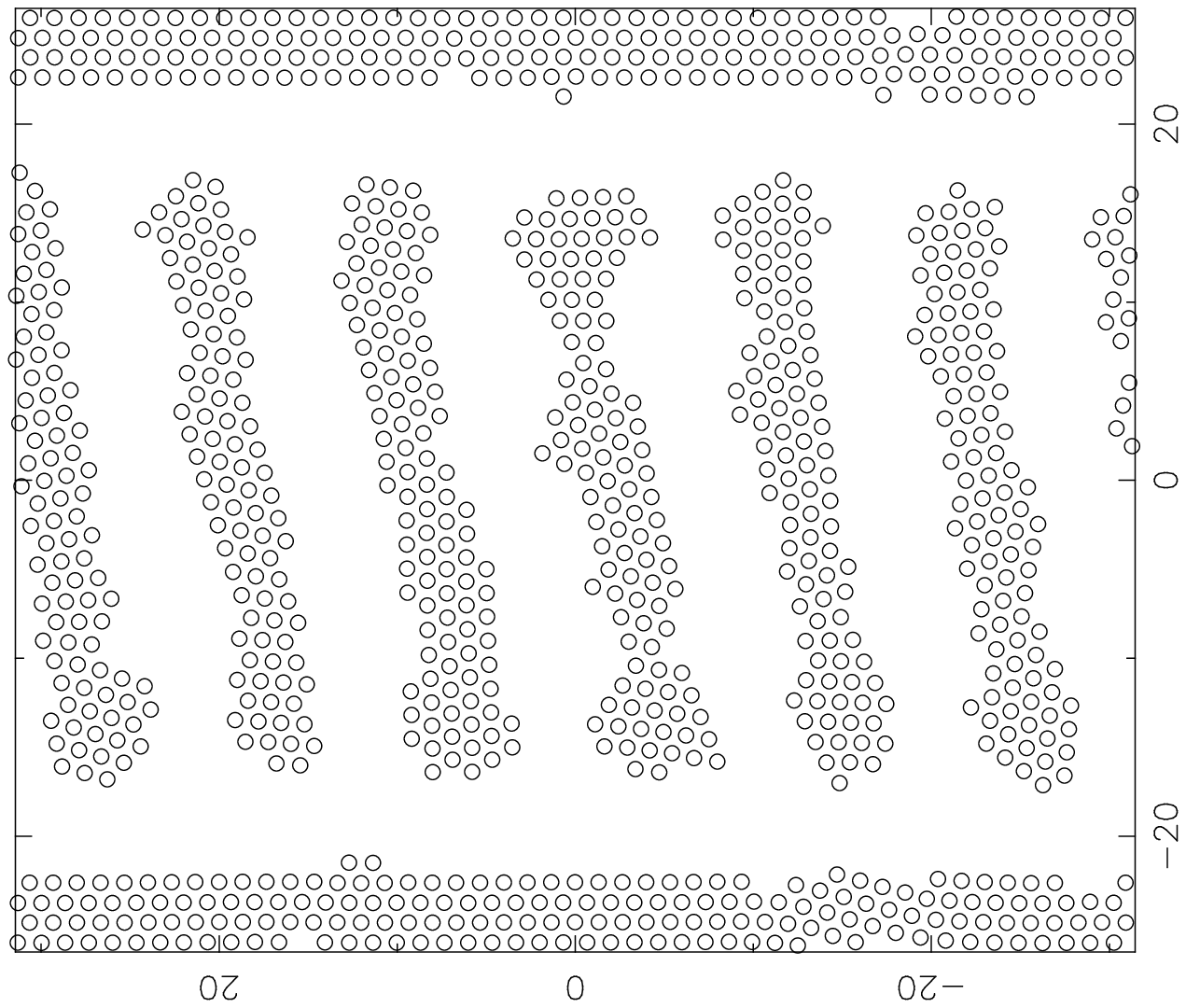}\includegraphics[width=3cm,angle=-90]{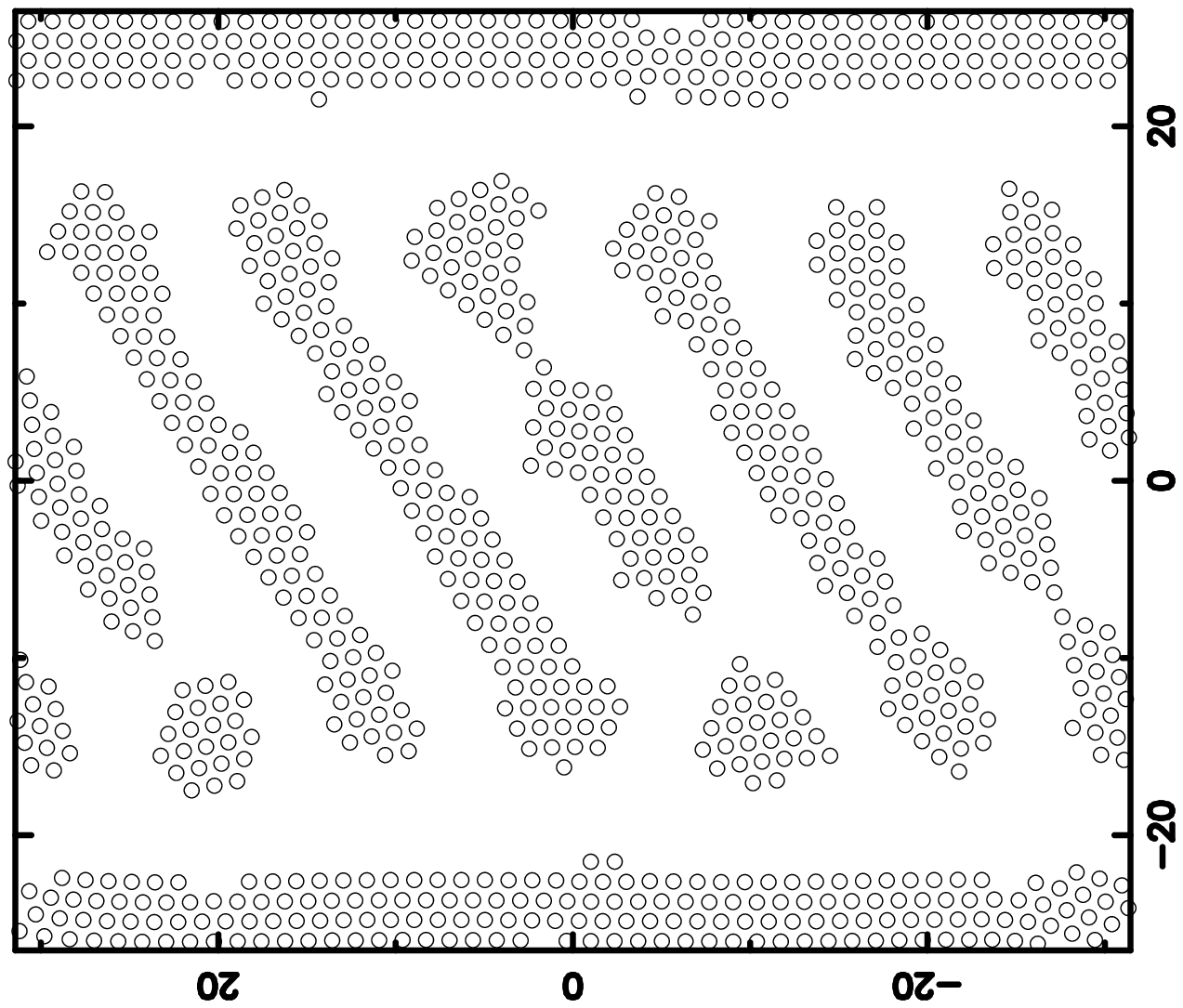}\includegraphics[width=3cm,angle=-90]{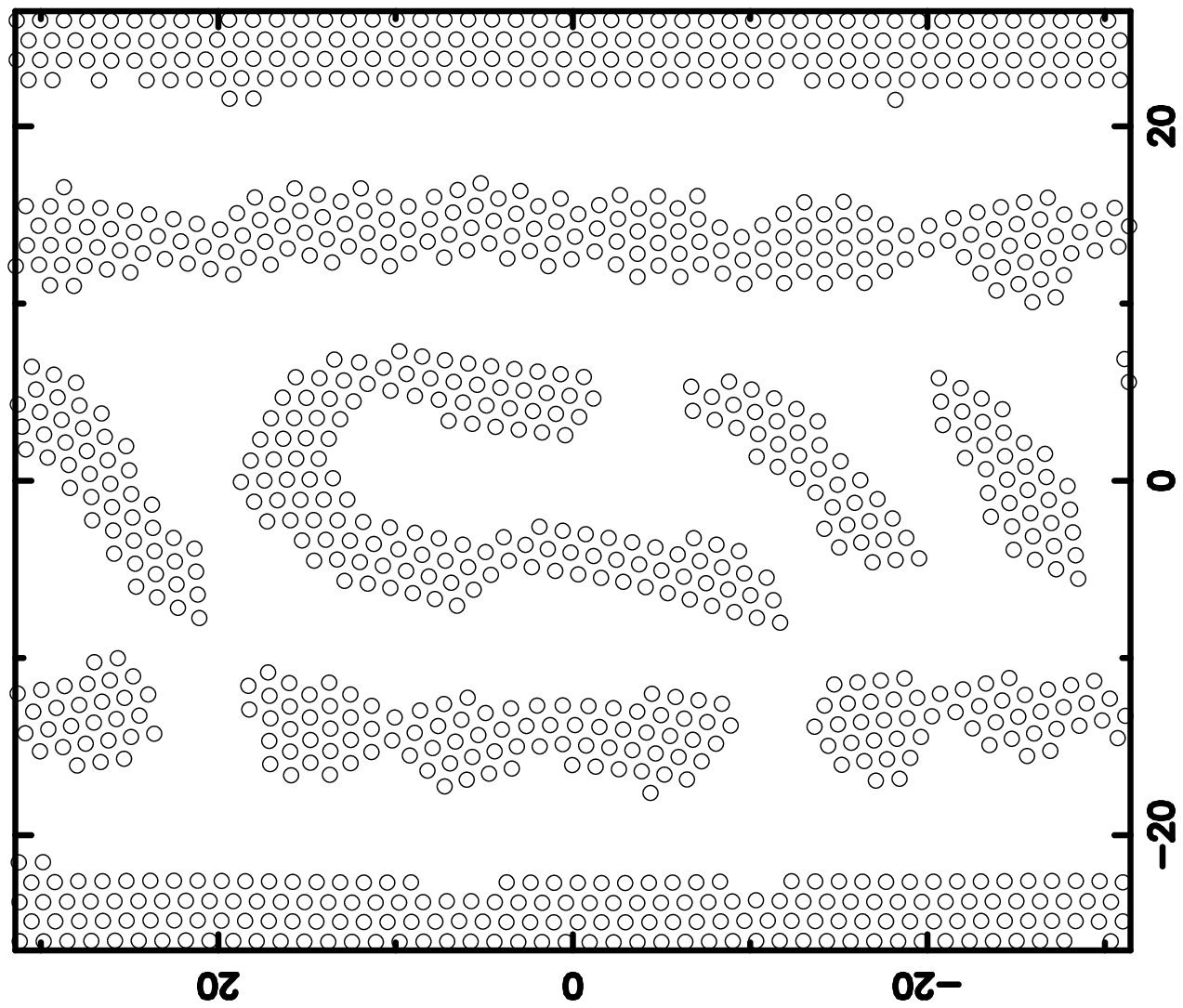}\\
\caption{\footnotesize{(Color online)
 Upper panel: the velocity profiles as a function of time for different values  of $F_0$, for
a system with $N=1000$, $\rho=0.3$ and $L_x =53$. At low shear rate only
the outer (parallel) stripes moves while the perpendicolar stripes are
jammed. At higher shear rate the perpendicolar stripes break and flow.
The dashed line represents the imposed profile. Bottom panels: particle configurations under different conditions. Initial configuration (left); yield point, when the 
central stripes break at $F=0.1$ and $\gamma=$ (middle); high strain $F_0=$ and $\gamma=1$ (right).   }}
\label{rho03prof_per}
\end{figure}

The jamming of the system is clearly visible in the  velocity profiles, where we see that the central part of the system has zero velocity and only the two boundary parallel stripes are flowing (Fig.~\ref{rho03prof_per}, upper panel). Typical snapshots of the system, when it is jammed and when it is flowing are plotted in the bottom panels of Fig.~\ref{rho03prof_per} too (See EPAPS Document No. XXXX.2 for an animation of the shear of the perpendicular stripe phase showing the 
reorientation of the stripes at the yield point).

\vspace{1cm}

\begin{figure}[hbtp]
\includegraphics[width=8cm]{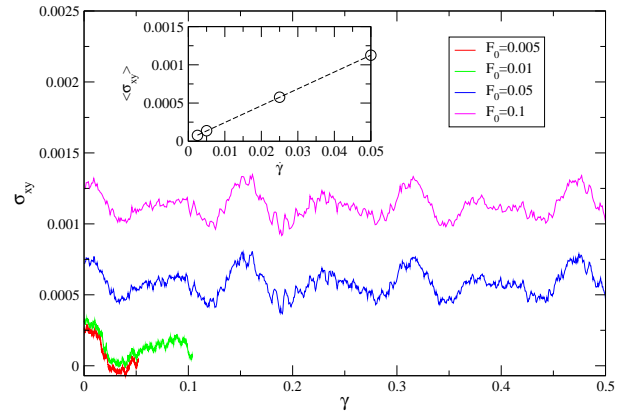}
\caption{\footnotesize{(Color online) The stress-strain curves for different values of $F_0$ for
a system with $N=1000$, $\rho=0.3$ and $L_x =48$. The initial state is characterized by stripes which are all parallel to the walls and to the flow direction.}}
\label{rho03stress_par}
\end{figure}

The case of parallel stripes is particularly simple to shear. The velocity profile rescaled to the applied force is always linear for each value of $F_0$ we have used. The most interesting aspect is perhaps the behavior of the stress strain-curves which 
shows some fluctuations, even  if much more smaller than those observed with perpendicular stripes, and probably due to the irregularity in the stripe surfaces
(see Fig.~\ref{rho03stress_par}). The mean stress is a linear function
of the strain rate, indicative of a simple viscous behavior. An extrapolation
at low shear rate would indicate a very low yield stress of the order of 
$\sigma_Y \simeq 10^{-5}$.

\clearpage

\section{Bubble phases \label{Bubble}}

\begin{figure}[hbtp]
\includegraphics[width=8cm]{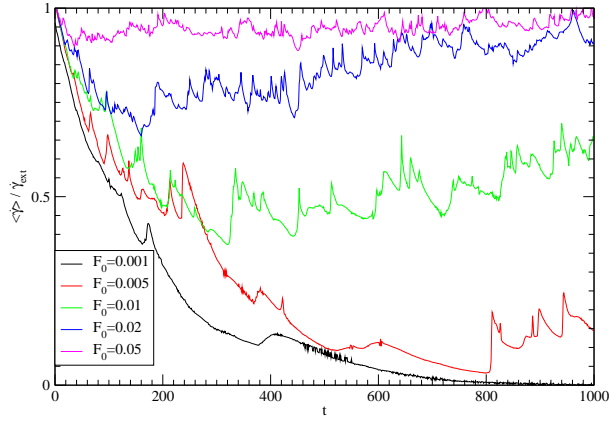}\\
\vspace{1cm}
\includegraphics[width=8cm]{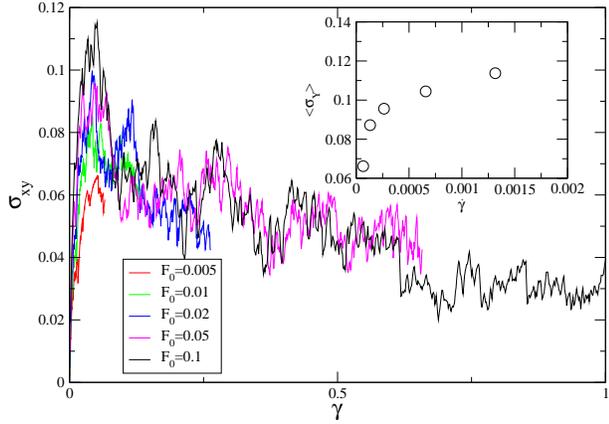}
\caption{\footnotesize{(Color online)Upper panel: the average strain rate as a function of time for different
values of $F_0$ for a system with $N~=~2000$, $\rho\sigma^2~=0.5$ and $L_x~=~76~\sigma$ plotted in log-linear scale. Notice that at large $F_0$ the strain rate reaches a steady value,
while for $F_0=0.0001$ the strain rate decreases towards zero, indicating a
jammed phase. Bottom panel:  the stress-strain curves for different values of $F_0$. In the inset we show the peak
stress as a function of the applied strain rate $\dot\gamma$.}}
\label{rho05gdot}
\end{figure}

\begin{figure}[hbtp]
\includegraphics[width=8cm]{rho05prof.eps}\\
\includegraphics[width=2cm,angle=-90]{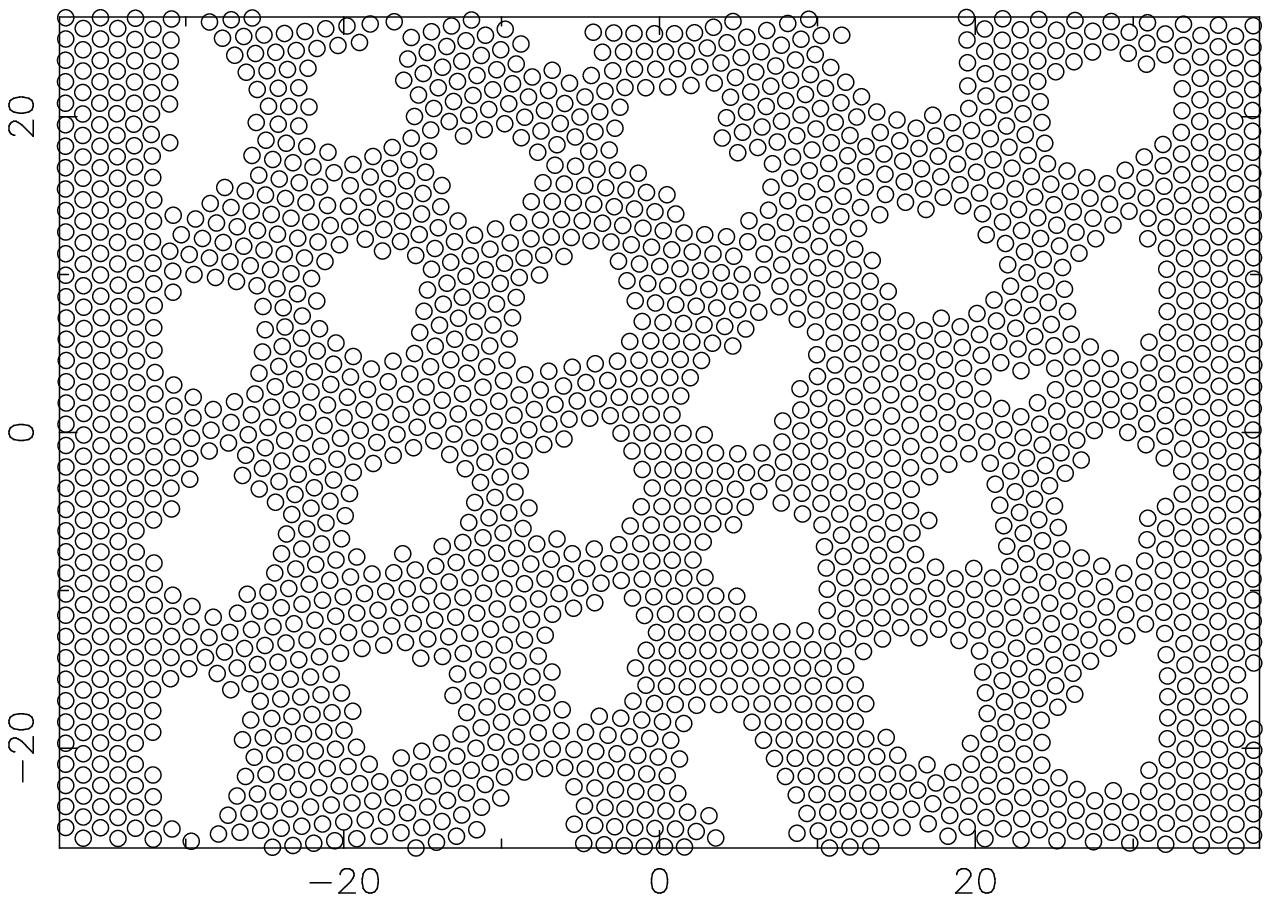}\includegraphics[width=2cm,angle=-90]{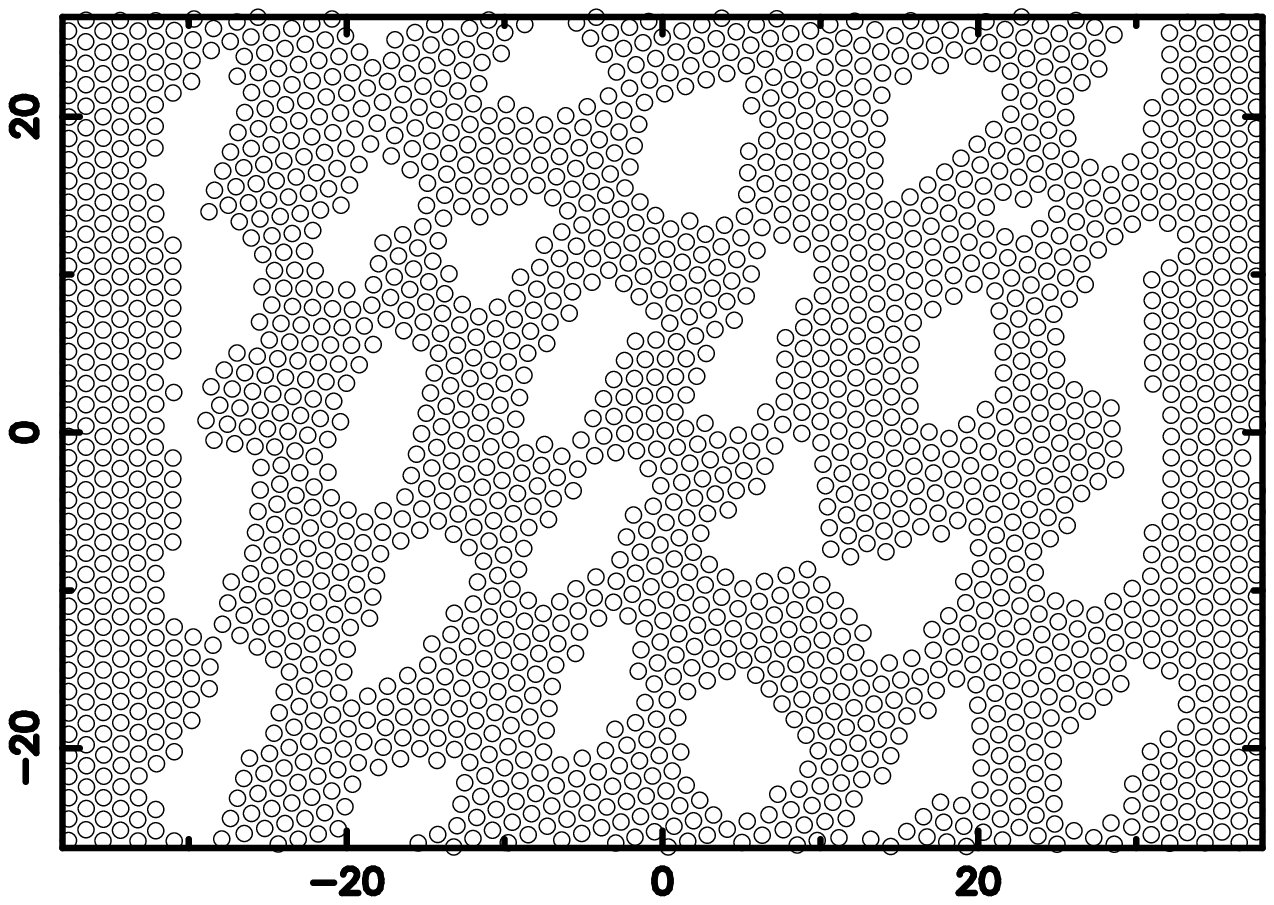}\includegraphics[width=2cm,angle=-90]{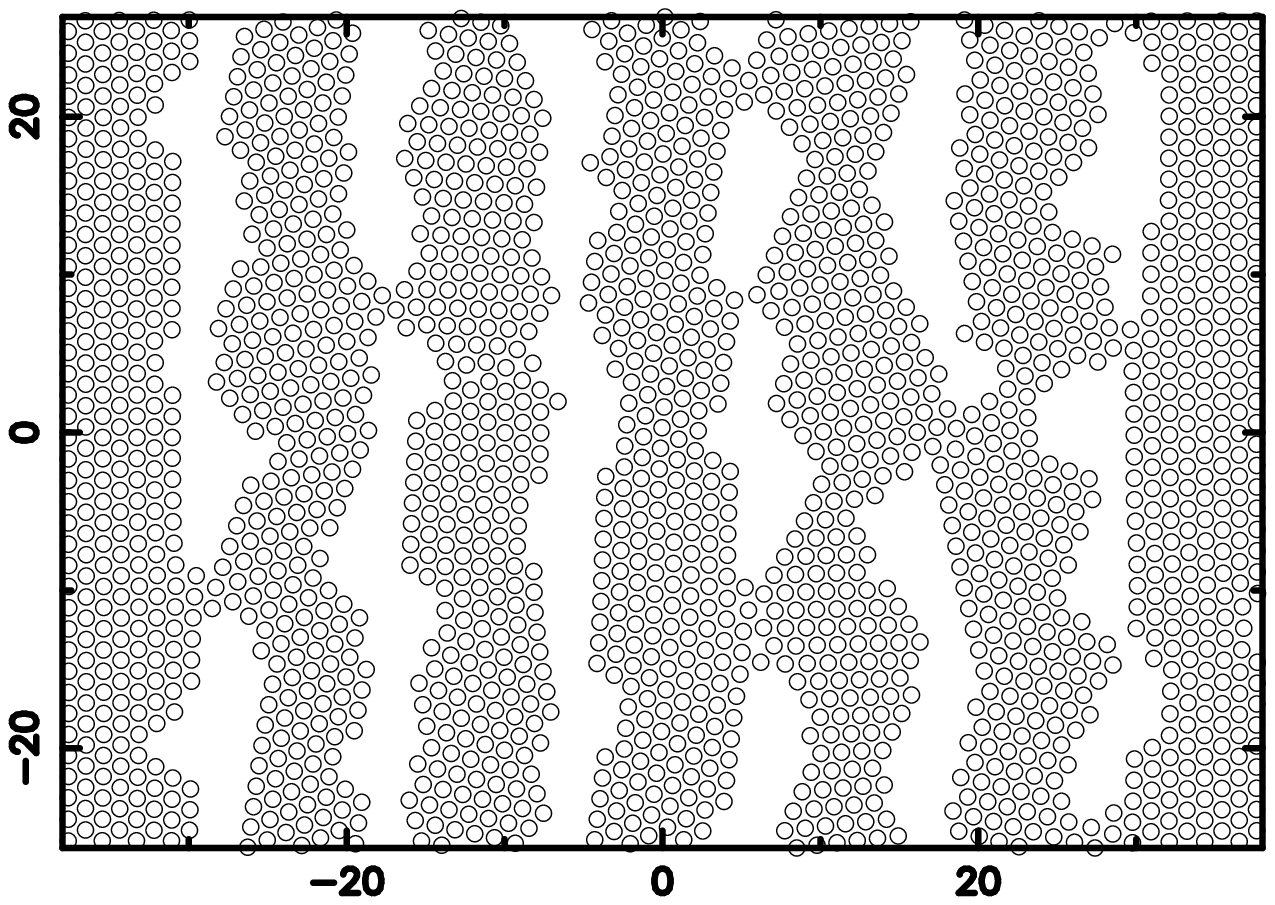}
\caption{\footnotesize{(Color online) Upper panel: the velocity profiles for different values of the drag force $F_0$ for
a system with $N~=~2000$, $\rho\sigma^2~=~0.5$ and $L_x~=~76\sigma$. Data are averaged over time
and over 10 realizations. Notice the deviation from the externally imposed 
linear profile occurring at lower shear rates.
For $F_0=0.0001$, the system is jammed and the profile is flat. Bottom panels: subsequent snapshots for the system at $F_0=0.1$. The snapshots are taken at the beginning of the simulation (left),
at the yield strain (middle) and at large strain (right). Notice the crossover from bubble to stripes at large strains.}
}
\label{rho05prof}
\end{figure}

Increasing the density, the system develops a bubble phase in which the bubbles are arranged on a triangular superlattice. From a qualitatively point of view, the bubbles under shear become more and more elongated along the direction of the applied velocity field, until they join to form stripe-like configurations (bottom panel of Fig.\ref{rho05prof}). The rheology of the bubble phase has some similarities with the one of the
perpendicular stripe phase, since both phases jam at low shear rates.
The main difference with respect to the stripe phase is that  for small external applied strain, the bubble phase is completely jammed. This is evident from the slowing down of the average strain rate for $F_0\leq 0.005$ (Fig. \ref{rho05gdot} upper panel) and also from the velocity profile which is completely flat (Fig. \ref{rho05prof} upper panel). In the case of perpendicular stripes, instead, the jamming involves only the central region, while the stripes close to the walls flow as well. 

When the applied force to the bubble phase is greater than the peak stress, the system starts to flow and once again parallel stripes are formed (Fig. \ref{rho05gdot} bottom panel), since this state is the easiest to shear (See EPAPS Document No. XXXX.3 for an animation of the shear induced stripe formation in the bubble phase). The peak stress is an increasing function of the strain rate as depicted in the inset of the bottom part of Fig.~\ref{rho05gdot}. Extrapolations at low shear rate suggest an yield stress of the order $\sigma_Y= 0.02$. The shear weakening observed in the stress-strain curve
is reminiscent of the stripe phase, but the yield stress $\sigma_Y$ measured in the bubble phase is about an order of magnitude greater than that observed in the stripe case. This is easily understandable as, in the bubble phase, the particles are strongly packed together, so that the motion of the individual particles is hindered by the surronding ones. In other words, collective motion of particles is necessary to ensure a structural change of the system. We notice as well that the stress-strain curve is characterized by large fluctuations, suggesting again intermittent and abrupt rearrangements as commonly observed in plasticity \cite{uchic04,dimiduk06,miguel01,csikor07} and in foam rheology \cite{durian99,tewari99,okuzono94}.

\begin{figure}[hbtp]
\includegraphics[width=7cm]{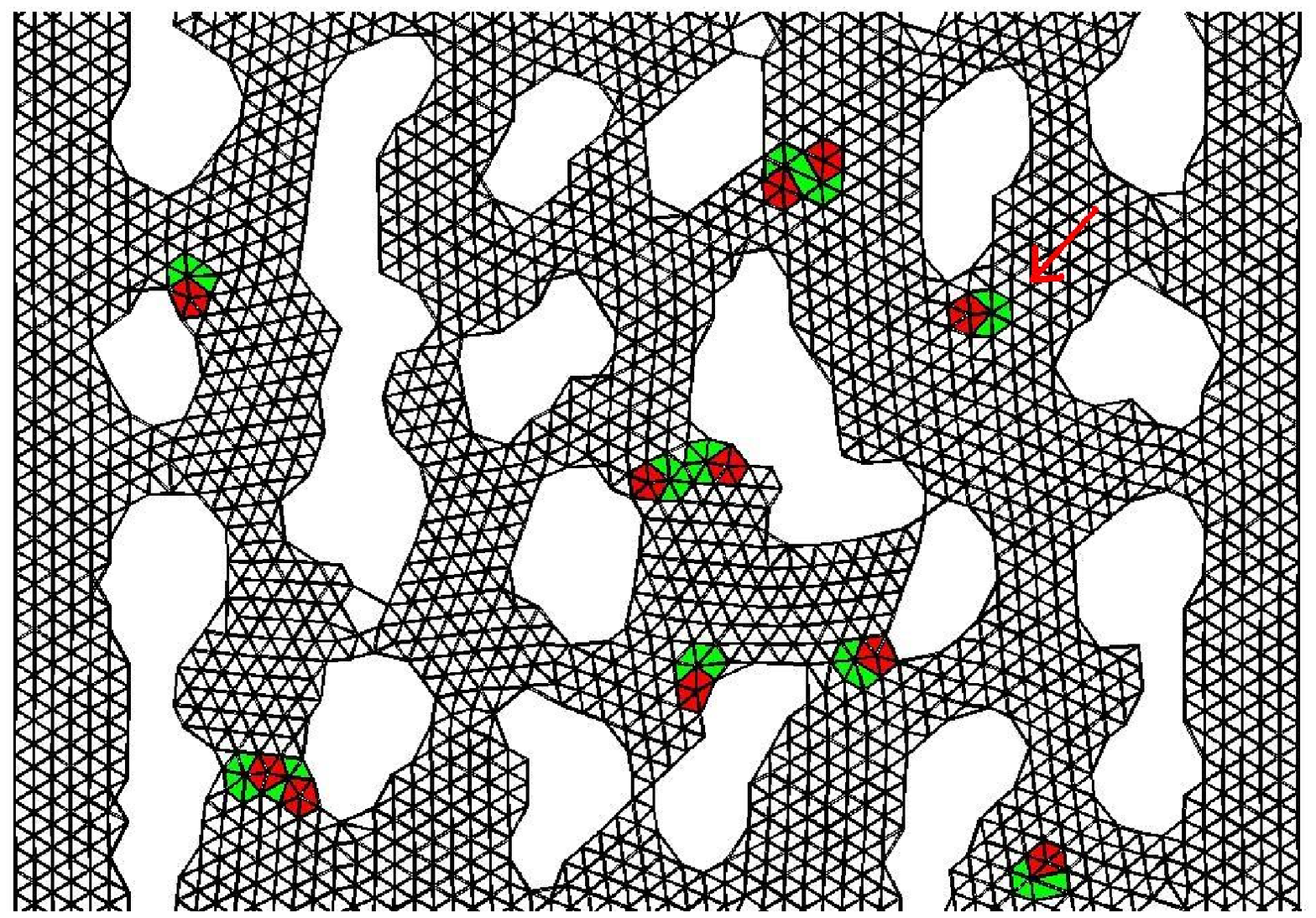}\\
\includegraphics[width=7cm]{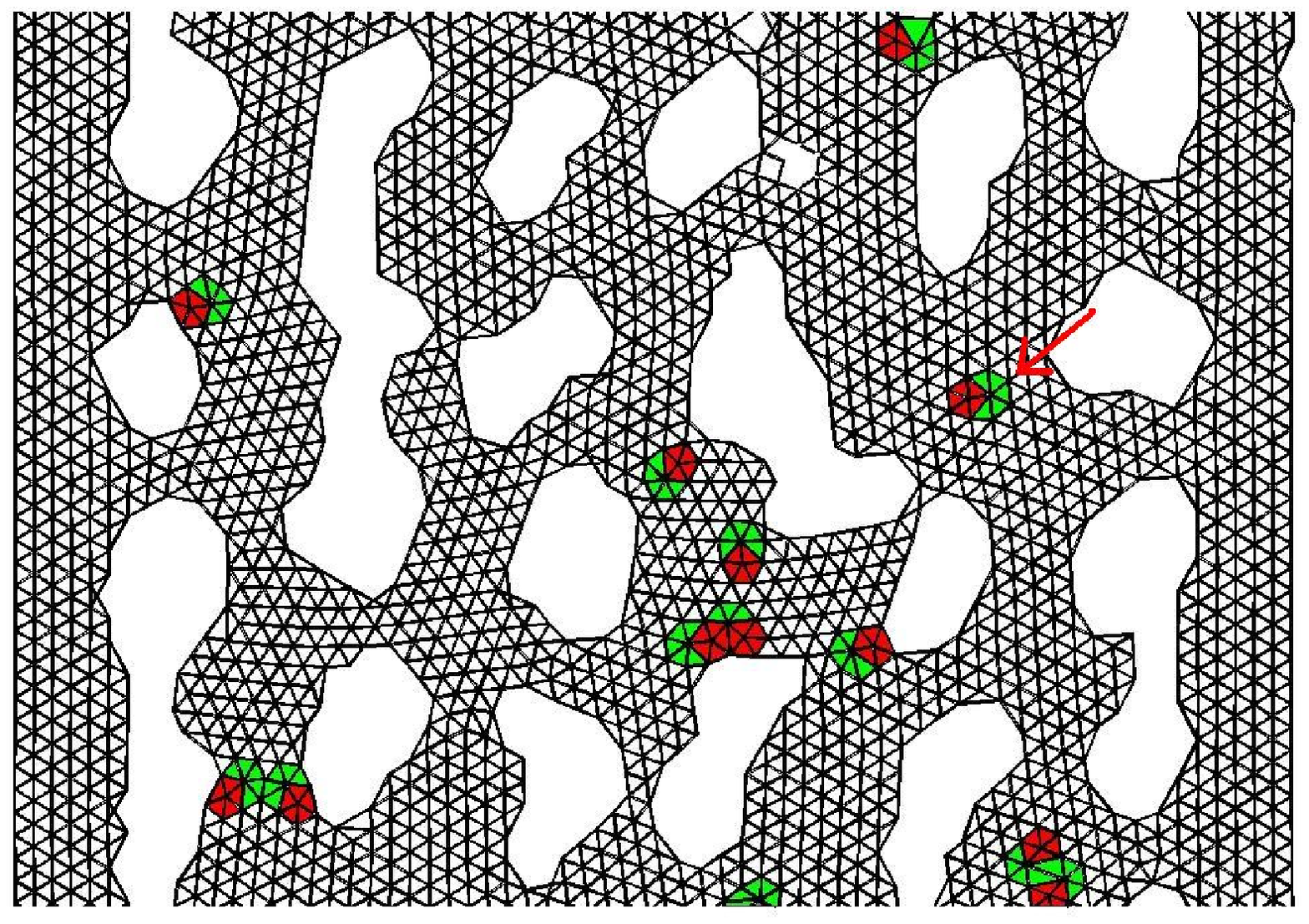}\\
\caption{(Color online) Voronoi triangulations of the configuration of the system configurations
at two subsequent time steps for a system with $N=2000$, $\rho=0.5$, $L_x =76$ and $F_0=0.1$. The snapshot are taken around the yield strain $\gamma\simeq 0.11$. 
Pairs of seven-fold and five-fold particles form dislocations and are colored respectively in green and red. For clarity, we do not show the topological defects arising at the sample boundary and around the bubbles. Notice the motion of a dislocation between neighboring bubbles (indicated by an arrow).}
\label{fig:dislocations}
\end{figure}

In order to better understand the yielding of the bubble phase, we perform a voronoi
triangulation of the particle system and follow the evolution of the topological defects.
Due to the $T=0$ condition, the system is locally ordered with a few topological defects around the bubbles.
As the system is sheared, we observe the nucleation and propagation of dislocations, corresponding to pairs of five-fold and seven-fold coordinated atoms (See EPAPS Document No. XXXX.4 for an animation of the voronoi triangulations of the configurations in the sheared bubble phase). I
n particular, dislocations are sometimes created at the surface
of a bubble and then propagate towards the nearest bubble (see Fig.~\ref{fig:dislocations}). Hence, we
can see the shear induced transformation from bubbles to stripes as a ductile fracture process: 
the system first deforms plastically in the vicinity of the bubbles and eventually a crack
propagates, leading to the coalescence of the bubbles into stripes.

It is also interesting to compare the rheology of the bubble phase with the one of the cluster phase.
At equilibrium, the cluster phase and the bubble are characterized by configurations which are specular. Under shear, however, the clusters are easily sheared also for very low values of $F_0$, while a yield stress must be overcome in the bubble phase. Moreover, while the clusters rotate and mantain their shape until $F_0$ is rather large, the bubbles deform almost immediately. Notwithstanding these large differences in the structure of the systems, the velocity profiles appear to be very similar for $F_0>0.005$. In particular the velocity profiles obtained in the bubble phase are similar to those obtained in the cluster phase but with smaller external forces. Moreover, the velocity profiles tend to become linear at first in the middle of the box simulation and eventually nearby the walls, indicating once again the presence of shear banding.

\vspace{1cm}

\clearpage

\section{Discussion and conclusions \label{Discussion}}

\begin{figure}[hbtp]
\includegraphics[width=8cm]{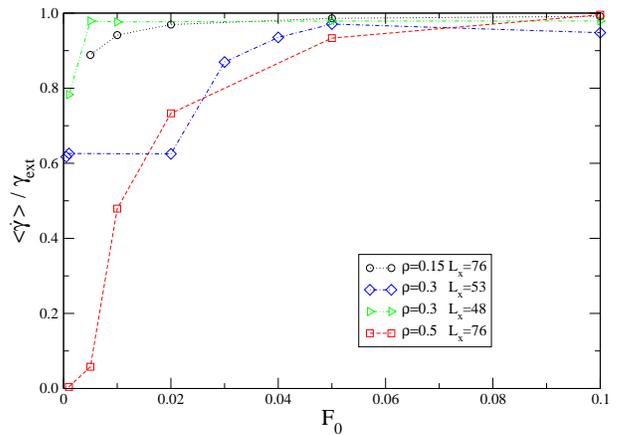}\\
\caption{(\footnotesize{Color online)The average strain rate, divided by the external
strain rate $\dot\gamma_{ext} \equiv V_0/(2L_x)$, as a function of the external force $F_0$ for different pattern morphology at equilibrium. The cluster case corresponds to $\rho=0.15$ $L_x=76$, the parallel stripes to $\rho=0.3$ $L_x=53$, the perpendicular stripes to $\rho=0.3$ $L_x=48$, the bubble case to $\rho=0.5$ $L_x=76$. }}
\label{summary}
\end{figure}

In simple liquids the flow properties are entirely determined by the viscosity which does not depend on the shear rate. In the laminar regime, for simple liquids a linear relation between $\sigma$ and $\dot\gamma$ holds. For the system we study, such behavior is not present anymore, because particle aggregates lead to a coupling between the microstructure of the system and its flow. Sometimes, the coupling between microstructure and flow is so strong that we obtain pattern morphologies which are not possible at equilibrium for a fixed density or for a fixed wall separation.\\
To summarize the behavior of the system at different densities, we compare the strain rate obtained for different pattern morphology. In particular, the strain rate is averaged over time and over 10 different simulations; it is also normalized to the external applied strain, in order to quantify the discrepancy with respect to a laminar flow. The results, as a function of the applied external force $F_0$, are plotted in Fig. \ref{summary}. 

For each pattern we see that, for sufficiently high values of $F_0$, the flow is laminar, that is the ratio $ <\dot\gamma>/\dot\gamma_{ext}$ tends to unity.
 For smaller values of the drag forces, the behavior of the systems is rather different and strictly dependent on the equilibrium configuration.
In particular the cluster phase is easy to shear. In this case
 the system is never jammed for the values of $F_0$ we have used, suggesting that if a yield stress exists, it is very small.
When the system is already arranged on stripes parallel to the walls, the flow is laminar even for the lowest values of $F_0$.
In the case of perpendicular stripes the strain rate profile is characterized at first by a plateau, which is connected to the coexistence of the flowing stripes nearby the walls and the jammed central stripes. Then at $F_0\geq 0.03$  the stripes breaks and tend to rearrange in the flow direction.
Finally in the bubble phase, the system is at first completely jammed as it demonstrated by the zero value of the ratio $ <\dot\gamma>/\dot\gamma_{ext}$. It seems however that in case of perpendicular stripes, the partial jamming persists on a wider range of $F_0$ with respect to the bubble phase.
Finally we observe that in the cases in which the system undergoes a structural change, such as that from the perpendicular alignment of stripes to the parallel one, or from the melting of the bubble phase to reform parallel stripes, a peak occurs in the strain-stress curve. The height of such a peak in the bubble phase is greater than the corresponding one in the stripe phase for an order of magnitude. Moreover strong fluctuations  appear in the stress-strain curve, which might suggest similarities with systems in which avalanche-like rearragements occur.

\vspace{2cm}

The main conclusion of our study is that for sufficiently strong shear rate the 
system dynamically forms stripes parallel to the flow direction. This fact is particularly evident at high densities ($\rho\sigma^2\geq 0.3$). In particular the parallel orientation appears to be preferred even when the geometrical constraints impose an equilibrium perpendicular orientation or when the equilibrium pattern has a completely different symmetry such as in the bubble phase at high density. 
The rheological behavior discussed here has been obtained at $T=0$. It is thus
important to discuss which features are expected to persist when thermal
effects are taken into account. We can expect that if the temperature is below
the melting transition, the behavior observed at high shear rates should be independent
on the temperature. At low shear rates, close to the jamming transition, thermal
fluctuations could help overcome the geometrical contraints leading to a slow
creep deformation. It would also be interesting to study the response of the model under the combined effect of thermal fluctuations and oscillating shear stresses.

\clearpage


\begin{thebibliography}{10}

\bibitem{klokkenburg06}
M.~Klokkenburg, R.P.A. Dullens, W.K. Kegel, B.H. Erne, and A.P. Philipse.
\newblock {\em Phys. Rev. Lett.}, 96:037205, 2006.

\bibitem{gelbart99}
W.M. Gelbart, R.P. Sear, J.R. Heath, and S.~Chaney.
\newblock {\em Faraday Discuss.}, 112:299, 1999.

\bibitem{elias97}
F.~Elias, C.~Flament, J.C. Bacri, and S.Neveu.
\newblock {\em J. Phys. I France}, 7:711, 1997.

\bibitem{ghezzi97}
F.~Ghezzi and J.C. Earnshaw.
\newblock {\em J. Phys.: Condens. Matter}, 9:L517, 1997.

\bibitem{seul95}
M.~Seul and D.~Andelman.
\newblock {\em Science}, 267:476, 1995.

\bibitem{bardi07}
F.~Bardi, C.~Cametti, and S.~Sennato.
\newblock {\em Colloid and surfaces A}, 306:102, 2007.

\bibitem{lu06}
P.J. Lu, J.C. Conrad, H.M. Wyss, A.B. Schofield, and D.A. Weitz.
\newblock {\em Phys. Rev. Lett.}, 96:028306, 2006.

\bibitem{stradner04}
A.~Stradner, H.~Sedwick, F.~Caedinaux, W.C.K. Poon, S.U. Egelhaaf, and
  P.~Schurtenberger.
\newblock {\em Nature}, 432:492, 2004.

\bibitem{islam03}
M.F. Islam, K.H. Lin, D.~Lacoste, T.C. Lubensky, and A.G. Yodh.
\newblock {\em Phys. Rev. E}, 67, 2003.

\bibitem{sciortino04}
Sciortino F., Mossa S, Zaccarelli E., and Tartaglia P.
\newblock {\em Phys. Rev. Lett.}, 93:055701, 2004.

\bibitem{charbonneau06}
P.~Carbonneau and D.R. Reichman.
\newblock {\em Phys. Rev. E}, 75(050401(R)), 2007.

\bibitem{decandia06}
A.~{de Candia}, E.~{Del Gado}, A.~Fierro, N.~Sator, M.~Tarzia, and A.~Coniglio.
\newblock {\em Phys. Rev. E}, 74:010403(R), 2006.

\bibitem{hoare07}
T.~Hoare and R.~Pelton.
\newblock {\em J. Phys. Chem. B}, 111:1334, 2007.

\bibitem{hecht07}
Hecht M., Harting J., and Herrmann H.J.
\newblock {\em Phys. Rev. E}, 75(5):051404, 2007.

\bibitem{reynaert06}
Reynaert S., Moldenaers P., and Vermont J.
\newblock {\em Langmuir}, 22:4936, 2006.

\bibitem{puertas05}
J.B. Caballero, A.M. Puertas, A.~Fernandez-Barbero, and F.~Javier de~las
  Nieves.
\newblock {\em Colloids and surfaces A}, 270-271:285, 2005.

\bibitem{campbell04}
A.I. Campbell, V.J. Anderson, P.~Bartlett, and J.S. van Duijneveldt.
\newblock {\em Phys. Rev. Lett.}, 94:208301, 2005.

\bibitem{destainville06}
N.~Destainville.
\newblock {\em Biophysical Journal - Biophysical Letters}, (L01), 2006.

\bibitem{goldman05}
J.~Goldman, S.~Andrews, and D.~Bray.
\newblock {\em Eur. Biophys. J}, 33:506, 2004.

\bibitem{szabo07}
A.~Szabo, E.D. Perryn, and A.~Czirok.
\newblock {\em Phys. Rev. Lett.}, 98:038102, 2007.

\bibitem{kern91}
K.K. Kern, H.~Niehus, A.~Schatz, P.~Zeppenfeld, J.~Goerge, and G.~Comsa.
\newblock {\em Physical Review Letters}, 67(7):855, 1991.

\bibitem{muller95}
Muller S., Bayer P, Reischl C., Heinz K., Feldmann B., Zillgen H, and Wuttig M.
\newblock {\em Phys. Rev. Lett.}, 74:765, 1995.

\bibitem{mladek06}
B.M. Mladek, G.~Kahl D.~Gottwald, M.~Neumann, and C.N. Likos.
\newblock {\em Phys. Rev. Lett.}, 96:045701, 2006.

\bibitem{glaser06}
M.A. Glaser, G.M. Grason, and R.D. Kamien.
\newblock Soft spheres make more mesophases.
\newblock {\em Europhys. Lett.}, 78:46004, 2007.

\bibitem{malescio03}
G.~Malescio and G.~Pellicane.
\newblock {\em Nature Materials}, 2:97, 2003.

\bibitem{barci07}
D.~Barci and D.~Stariolo.
\newblock {\em Phys. Rev. Lett.}, 98:200604, 2007.

\bibitem{tarzia07}
Tarzia M.~Coniglio A.
\newblock {\em Phys. Rev. E}, 75:011410, 2007.

\bibitem{pini00}
D.~Pini, G.~Jialin, A.~Parola, and L.~Reatto.
\newblock {\em Chemical Physics Letters}, 327:209, 2000.

\bibitem{reichhardt04}
C.~J.~Olson Reichhardt, C.~Reichhardt, and A.~R. Bishop.
\newblock {\em Phys. Rev. Lett.}, 92, 2004.

\bibitem{reichhardt03}
I.~Martin C.~Reichhardt, C. J.~Olson and A.~R. Bishop.
\newblock Depinning and dynamics of systems with competing interactions in
  quenched disorder.
\newblock {\em EPL (Europhysics Letters)}, 61(2):221, 2003.

\bibitem{imperio06}
Imperio A. and Reatto L.
\newblock {\em J. Chem. Phys.}, 124:164712, 2006.

\bibitem{stoycheva02}
A.D. Stoycheva and S.J. Singer.
\newblock {\em Phys. Rev. E}, 65:036706, 2002.

\bibitem{larson92}
R.G. Larson.
\newblock {\em J. Chem. Phys.}, 96(11):7904, 1992.

\bibitem{tsori01}
Y.~Tsori and D.~Andelman.
\newblock {\em Eur. Phys. J. E}, 5:605, 2001.

\bibitem{tsori06}
Y.~Tsori and D.~Andelman.
\newblock {\em Journal of polymer science, B}, 44:2725, 2006.

\bibitem{li06}
W.~Li, R.A. Wickham, and R.A. Garbary.
\newblock {\em Macromolecules}, 39:806, 2006.

\bibitem{duchs04}
D.~Duchs and F.~Schmidt.
\newblock {\em J. Chem. Phys.}, 121(4):2798, 2004.

\bibitem{impe07}
Imperio A. and Reatto L.
\newblock {\em Phys. Rev. E}, 76(040402(R)), 2007.

\bibitem{gopal95}
A.D. Gopal and D.J. Durien.
\newblock {\em Phys. Rev. Lett.}, 75:2610, 1995.

\bibitem{gopal05}
A.D. Gopal and D.J. Durien.
\newblock {\em Phys. Rev. Lett.}, 91(18):188303--1, 2005.

\bibitem{lootens05}
D.~Lootens, H.~van Damme, Y.~Hemar, and P.~Hebrand.
\newblock {\em Phys. Rev. Lett.}, 95:268302, 2005.

\bibitem{osuji07}
C.O. Osuji, C.~Kim, and D.A. Weitz.
\newblock {\em arXiv:0710.0042}, 2007.

\bibitem{cohen05}
I.~Cohen, T.G. Mason, and D.~Weitz.
\newblock {\em Phys. Rev. Lett.}, 93(4):046001, 2005.

\bibitem{melrose92}
J.R. Melrose and D.M. Heyes.
\newblock {\em J. Chem. Phys.}, 98(7):5873, 1992.

\bibitem{trappe01}
V.~Trappe, V.~Prasad, L.~Cipelletti, P.N. segre, and D.A. Weitz.
\newblock {\em Nature}, 411:772, 2001.

\bibitem{walder07}
R.~Walder, C.~Schmidt, and M.~Dennin.
\newblock {\em arXiv:0711.1628}, 2007.

\bibitem{ignes-mullol07}
J.~Ignes-Mullol, J.~Claret, R.~Reigada, and F.~Sagues.
\newblock {\em Phys. Rep.}, 448:163, 2007.

\bibitem{lauridsen05}
J.~Lauridsen, G.~Chanan, and M.~Dennin.
\newblock {\em Phys. Rev. Lett.}, 93(1):018303, 2005.

\bibitem{durian99}
D.J. Durian.
\newblock {\em Phys. Rev. E}, 55(2):1739, 1999.

\bibitem{tewari99}
S.~TGewari, D.~Schiemann, D.J. Durian, C.M. Knobler, S.A. Langer, and A.J. Liu.
\newblock {\em Phys. Rev. E}, 60(4):4385, 1999.

\bibitem{okuzono94}
T.~Okuzono and K.~Kawasaki.
\newblock {\em Phys. Rev. E}, (2):1246, 1995.

\bibitem{maloney04}
Craig Maloney and Ana\"{e}l Lema\^{i}tre.
\newblock {\em Physical Review Letters}, 93(1):016001, 2004.

\bibitem{demkowicz05}
Michael~J. Demkowicz and Ali~S. Argon.
\newblock {\em Physical Review B (Condensed Matter and Materials Physics)},
  72(24):245206, 2005.

\bibitem{bailey07}
Nicholas~P. Bailey, Jakob~Schi\o tz, Ana\"{e}l Lema\^{i}tre, and Karsten~W.
  Jacobsen.
\newblock {\em Physical Review Letters}, 98(9):095501, 2007.

\bibitem{uchic04}
M.~D. {Uchic}, D.~M. {Dimiduk}, J.~N. {Florando}, and W.~D. {Nix}.
\newblock {\em Science}, 305:986--989, aug 2004.

\bibitem{dimiduk06}
D.~M. {Dimiduk}, C.~{Woodward}, R.~{LeSar}, and M.~D. {Uchic}.
\newblock {\em Science}, 312:1188--1190, may 2006.

\bibitem{miguel01}
M.-C. {Miguel}, A.~{Vespignani}, S.~{Zapperi}, J.~{Weiss}, and J.-R. {Grasso}.
\newblock {\em Nature}, 410:667--671, apr 2001.

\bibitem{csikor07}
F.~Csikor, C.~Motz, D.~Weygand, M.~Zaiser, and S.~Zapperi.
\newblock {\em Science}, 318:251, 2007.

\bibitem{astrom00}
J.A. Astrom, H.J. Herrmann, and J.~Timonen.
\newblock {\em Phys. Rev. Lett.}, 84(4):638, 2000.

\bibitem{tsamados07}
M.~Tsamados, A.~Tanguy, F.~Leonforte, and J.L. Barrat.
\newblock {\em arXiv:0711.3127}, 2007.

\end{thebibliography}
\end{document}